\documentclass[journal]{IEEEtran}
\usepackage[utf8]{inputenc}
\usepackage{cite}

\usepackage{svg}
\usepackage{amsmath}
\usepackage{amssymb}
\usepackage{hyperref}
\usepackage{relsize}%
\usepackage{graphicx}
\usepackage{subfig}
\usepackage{xcolor}
\usepackage[english]{babel}
\usepackage{amsthm}
\usepackage{xurl}
\newtheorem{remark}{Remark}
\usepackage{algorithmicx}
\usepackage[ruled,vlined, linesnumbered]{algorithm2e}
\usepackage{comment} 
\newcommand{\ignore}[1]{}
\usepackage{color} 
\usepackage{multirow}
\usepackage[T1]{fontenc}
\usepackage{lettrine}

\makeatletter
\newcommand{\removelatexerror}{\let\@latex@error\@gobble}

\IEEEoverridecommandlockouts
\begin{document}

\title{\LARGE \bf Communication-Efficient Decentralized Multi-Agent Reinforcement Learning for Cooperative Adaptive Cruise Control}

\author{Dong Chen$^{1}$, \textit{Member, IEEE}, Kaixiang Zhang$^{2}$, Yongqiang Wang$^{3}$, \textit{Senior Member, IEEE},  Xunyuan Yin$^{4*}$, Zhaojian Li$^{2*}$, \textit{Senior Member, IEEE}, Dimitar Filev$^{5}$, \textit{Life Fellow, IEEE}

\thanks{$^1$Dong Chen is with the Department of Electrical and Computer Engineering, Michigan State University, Lansing, MI, 48824, USA. Email: {\tt {chendon9}@msu.edu.}}
\thanks{$^2$Kaixiang Zhang and Zhaojian Li are with the Department of Mechanical Engineering, Michigan State University, Lansing, MI, 48824, USA. Email: {\tt\{zhangk64, lizhaoj1\}@msu.edu.}}
\thanks{$^3$Yongqiang Wang is with the Department of Electrical and Computer Engineering, Clemson University, Clemson, SC, 29630, USA. Email: {\tt{yongqiw@clemson.edu}}.}
\thanks{$^4$Xunyuan Yin is with the School of Chemistry, Chemical Engineering and Biotechnology, Nanyang Technological University, 62 Nanyang Drive, 637459, Singapore. Email:  {\tt xunyuan.yin@ntu.edu.sg}.}
\thanks{$^5$Dimitar Filev is with Hagler Institue for Advanced Study, Texas A\&M University, College Station, Texas, 77843-3572, USA. Email:  {\tt dimitar.filev@gmail.com}.
}

\thanks{$*$Zhaojian Li and Xunyuan Yin are corresponding authors.}

\thanks{This work was partially supported by the National Science Foundation projects CNS-2219488 and CNS-2219487.}}

\maketitle


\begin{abstract}
Connected and autonomous vehicles (CAVs) promise next-gen transportation systems with enhanced safety, energy efficiency, and sustainability. One typical control strategy for CAVs is the so-called cooperative adaptive cruise control (CACC) where vehicles drive in platoons and cooperate to achieve safe and efficient transportation. In this study, we formulate CACC as a multi-agent reinforcement learning (MARL) problem. Diverging from existing MARL methods that use centralized training and decentralized execution which require not only a centralized communication mechanism but also dense inter-agent communication during training and online adaptation, we propose a fully decentralized MARL framework for enhanced efficiency and scalability. In addition, a quantization-based communication scheme is proposed to reduce the communication overhead without significantly degrading the control performance. This is achieved by employing randomized rounding numbers to quantize each piece of communicated information and only communicating non-zero components after quantization. Extensive experimentation in two distinct CACC settings reveals that the proposed MARL framework consistently achieves superior performance over several contemporary benchmarks in terms of both communication efficiency and control efficacy. In the appendix, we show that our proposed framework's applicability extends beyond CACC, showing promise for broader intelligent transportation systems with intricate action and state spaces.

\end{abstract}

\IEEEpeerreviewmaketitle

\section{Introduction}\label{sec:1}
\lettrine{C}onnected and autonomous vehicles (CAVs) have recently gained significant attention due to their promise to create safe and sustainable future transportation systems \cite{SOS, han2023secure, Planning, cao2022future, zhang2023emerging}. One pivotal technology of CAVs, known as cooperative adaptive cruise control (CACC), has been recognized for its capability to increase road efficiency, alleviate traffic congestion, and reduce both energy consumption and exhaust emissions \cite{kazemi2018learning, wu2023review, bolduc2019multimodel}. Specifically, by utilizing real-time vehicle-to-vehicle (V2V) communication, the primary objective of CACC is to adaptively coordinate a fleet of vehicles as a means to minimize the car-following headway and speed variations while preserving safety \cite{wen2018cooperative, lin2020comparison, wang2020online}. 

Yet, developing a robust CACC paradigm that tightly integrates computing, communication, and control technologies presents a considerable challenge, especially in the presence of limited onboard communication bandwidth and constrained computing resources \cite{hu2023cacc, althoff2020provably}. Classical control theory and optimization-based methodologies have been employed to tackle the CACC problem \cite{massera2017safe, gao2016data, al2017feedforward, wu2018stabilizing, gong2019cooperative}. Specifically, some research targets the car-following model \cite{massera2017safe} and string stability \cite{wang2018infrastructure, feng2019string}, modeling CACC within the context of a two-vehicle system. In contrast, other studies pose the CACC as optimal control problems \cite{jin2014dynamics, gao2016data, wu2018stabilizing}. These approaches hinge on precise system modeling \cite{massera2017safe, wang2018infrastructure, feng2019string} that are not generally available. They also typically involve online optimization, which requires significant computation resources to support real-time engineering systems \cite{chu2019model}. 

On the other hand, CAV platoon control has also been conceptualized as a sequential decision-making problem and addressed using data-driven strategies such as reinforcement learning (RL) \cite{chu2019model, lei2022deep, jiang2022reinforcement, zhu2022joint, wang2022design, liu2022autonomous, li2023anti}. In particular, in \cite{jiang2022reinforcement},  Soft Actor-Critic (SAC) \cite{haarnoja2018soft} is adopted to mitigate traffic oscillations and enhance platoon stability. Furthermore, the deep deterministic policy gradient (DDPG) algorithm \cite{lillicrap2015continuous} is employed in \cite{wang2022design} for CACC, taking into account both time-varying leading vehicle velocity and communication delays via wireless V2V communication technology. A policy-gradient RL approach is developed in \cite{desjardins2011cooperative} to ensure a safe longitudinal distance to a front vehicle. However, these approaches primarily focus on a vehicle fleet of only 2 vehicles (i.e., leader-follower architecture). To control multiple CAVs, centralized RL approaches are commonly developed, which rely heavily on the high-bandwidth capabilities of vehicle-to-cloud (V2C) or vehicle-to-infrastructure (V2I) communication \cite{jia2016enhanced}. For instance, in \cite{chu2019model}, a centralized RL controller is introduced for the CACC problem in mixed-traffic scenarios via V2C communication. While these centralized control strategies have demonstrated promising results, they bear the burden of heavy communication overheads and are often plagued by a single point of failure and the curse of dimensionality \cite{chu2020multiagent}. These factors make them impractical for deployment in large-scale CACC systems prevalently envisioned in the future landscape.

More recently, multi-agent reinforcement learning (MARL) has emerged as a promising solution to address the CACC control problem involving multiple CAVs, owing to its capabilities of online adaptation and solving complex problems \cite{peake2020multi, chu2020multiagent, raja2022blockchain}. For instance, a MARL framework with both local and global reward designs is developed and evaluated in \cite{peake2020multi} on two platoons of 3 and 5 CAVs, concluding that the local reward design (i.e., independent MARL) outperforms the global reward design. However, our experiments demonstrate that while independent MARL achieves promising performance in straightforward CACC scenarios, it falls short in more complex situations (see Section~\ref{sec:5}). In \cite{hua2023energy}, the authors extend CACC strategies with a novel MADDPG \cite{lowe2017multi} algorithm that addresses energy management challenges through a relevance ratio to ensure cooperative agent behavior. In \cite{chu2020multiagent}, a learnable communication MARL protocol is developed to reduce information loss across two CACC scenarios, and each agent (i.e., AV) learns a decentralized control policy based on local observations and messages from connected neighbors. Moreover, blockchain is incorporated into the MARL (i.e., MADDPG) framework to enhance the privacy of CACC. Despite these advances, the aforementioned approaches uniformly adopt a \textit{Centralized Training and Decentralized Execution} (CTDE) strategy, wherein agents use additional global information to guide training in a centralized manner and make decisions based on decentralized local policies \cite{zhang2018fully, zhang2021multi}. However, in many real-world scenarios, such as CACC, deploying a central controller (e.g., cloud facilities or roadside units) for training or online adaptation can be prohibitively expensive and complex. Moreover, the central controller needs to communicate with all local agents to exchange information, which perpetually amplifies the communication overhead on the single controller \cite{zhang2018fully}.

In this paper, we formulate CACC as a fully decentralized MARL problem, in which the agents are connected via a sparse communication network without the need for a centralized controller. To achieve this, we introduce a decentralized MARL algorithm based on a novel policy gradient update mechanism. Throughout the training process, each agent takes an individual action based solely on locally available information at each step. To stabilize training and counteract the inherent non-stationarity in MARL \cite{zhang2021multi}, each agent shares its estimate of the value function with its neighbors on the network, collectively aiming to maximize the average rewards of all agents across the network. Furthermore, a novel quantization-based communication scheme is further proposed, which greatly improves the communication efficiency in decentralized stochastic optimization without a substantial compromise on optimization accuracy. The main contributions and the technical advancements of this paper are summarized as follows.

\begin{enumerate}
\item We formulate the CACC problem as a fully decentralized MARL framework, which facilitates fast convergence without relying on a centralized controller for both training and execution. The developed codes are available in our open-source repository\footnote{\url{https://github.com/DongChen06/MACACC}}.

\item We introduce a novel effective and scalable MARL algorithm, featuring a quantization-based communication protocol to significantly enhance communication efficiency without major performance compromise. The quantization process condenses complex parameters of the critic network into discrete representations, facilitating efficient information exchange among agents. 

\item We conduct comprehensive experiments on two CACC scenarios, and the results show that the proposed approach consistently outperforms several state-of-the-art MARL algorithms.

\item In the appendix, we show that our proposed framework's applicability extends beyond CACC, showing promise for broader intelligent transportation systems characterized by intricate action and state spaces.
\end{enumerate}

The remainder of this paper is organized as follows. Section~\ref{sec:2} provides a brief overview of RL and MARL concepts. In Section~\ref{sec:3}, the considered CACC problem is formulated. The problem formulation and the proposed MARL framework are introduced in Section~\ref{sec:4} whereas experiments, results, and discussions are presented in Section~\ref{sec:5}. Lastly, in Section~\ref{sec:6}, we conclude the paper, summarize our contributions, and suggest potential insights for future research.

\section{Background}\label{sec:2}
In this section, we provide an overview of the preliminaries of RL and several leading-edge MARL algorithms. These MARL algorithms will later serve as benchmarks for comparison in Section~\ref{sec:5}.

\vspace{-10pt}
\subsection{Preliminaries of Reinforcement Learning (RL)}
RL, often mathematically formulated as a Markov decision process (MDP), has shown great promise as a data-driven method for learning adaptive control policies \cite{chu2020multiagent}. Recent advancements in deep neural networks (DNNs) have further amplified their learning capabilities for intricate tasks. Successful examples of these algorithms include deep Q-network (DQN) \cite{mnih2015human}, deep deterministic policy gradient (DDPG) \cite{lillicrap2015continuous}, and advantage actor-critic (A2C \cite{mnih2016asynchronous}). 

In an RL setting, at each time step $t$, the agent observes the state $s_t \in \mathcal{S} \subseteq \mathcal{R}^n$ from the environment, and performs an action $a_t \in \mathcal{A} \subseteq \mathcal{R}^m$ according a learned policy $\pi(a_t | s_t)$. Then the environment evolves to a new state $s_{t+1}$ according to the transition dynamics $p(\cdot|s_t, a_t)$, and emits an immediate reward $r_t=r(s_t, a_t, s_{t+1})$ to the agent. The objective of an RL agent is to learn an optimal policy $\pi^*: \mathcal{S} \rightarrow \mathcal{A}$ that maps from state to action, maximizing the accumulated reward $R_t = \sum_{k=0}^{T} \gamma^k r_{t+k}$, where $r_{t+k}$ is the reward at time step $t + k$, and $\gamma \in (0, 1]$  and $T$ represent the discount factor and episode length, respectively. The state-action function is denoted as $Q^{\pi}(s_t, a_t)= \mathop{\mathbb{E}}(R_t|s_t, a_t)$, representing the expected return starting from state $s_t$ and taking an immediate action $a_t$, then following policy $\pi$ afterward. The optimal Q-function $Q^*(s_t, a_t) = \max_{\pi}Q^{\pi}(s_t, a_t)$ determines the optimal greedy policy $\pi^*(a_t|s_t)$. The state value function $V^{\pi} (s_t) = \mathop{\mathbb{E}}(R_t|s_t)$ represents the expected return if starting from $s_t$ and immediately following the policy $\pi$.

In Q-learning, the Q-function, denoted as $Q_{\theta}$, is usually parameterized by a set of parameters $\theta$, utilizing function approximators such as Q-tables \cite{watkins1992q}, linear regression (LR) \cite{szepesvari2022algorithms}, or DNNs \cite{mnih2015human}. The temporal difference $(\mathcal{T}Q_{\theta^-} - Q_{\theta})(s_t, a_t)$ is employed to update $\theta$, where $\mathcal{T}$ and $Q_{\theta^-}$ represent the dynamic programming (DP) operator and a frozen recent model $\theta^-$ \cite{chu2019model}, respectively. $\epsilon-\text{greedy}$ and experience replay are commonly applied in deep Q-learning to reduce the estimation variance \cite{szepesvari2022algorithms}. In contrast, the policy $\pi_\theta$ is typically directly approximated by a set of parameters $\theta$ within the policy gradient method. The update of $\theta$ aims to enhance the likelihood and the loss function, represented as $\mathcal{L} (\pi_{\theta}) =\mathop{\mathbb{E}}_{\tau \sim \pi_{\theta}}[\sum_{t=0}^T \nabla_{\theta} \log \pi_{\theta} (a_t|s_t) R_t]$. Compared to Q-learning, the policy gradient is robust to nonstationary transitions within each trajectory, despite suffering from high variance \cite{chu2019multi}. \cite{chu2019multi}. Actor-critic algorithms, such as A2C \cite{mnih2016asynchronous}, enhance the policy gradient method by introducing the advantage function $A^\pi (s_t, a_t) = Q^{\pi_\theta}(s_t, a_t) - V_w(s_t)$, thereby reducing the variance of sample return. The parameters $\theta$ are updated with the policy loss function, defined as $\mathcal{L}=\mathop{\mathbb{E}}_{\pi_{\theta}}[\sum_{t=0}^T \nabla_{\theta} \log \pi_{\theta} (a_t|s_t) A_t]$, while the value function is updated as $\mathcal{L}=\min_{w} \mathop{\mathbb{E}}_{\mathcal{D}}[(R_t + \gamma V_{w^-}(s_t) - V_w(s_t))^2]$, where $\mathcal{D}$ and $w^-$ represent the experience replay buffer accumulating previous experiences and the parameters from prior iterations used in a target network, respectively \cite{chen2023deep}. Nevertheless, RL often encounters scalability issues in numerous real-world control problems involving multiple controllable agents, attributed to non-stationarity and partial observability \cite{chu2020multiagent}.

\vspace{-10pt}
\subsection{Multi-Agent Reinforcement Learning (MARL)} \label{sec:2marl}
To tackle the challenges of scalability inherent in RL, MARL has been proposed, in which each individual agent can adapt and learn its specific policy based solely on its local observations \cite{chu2019multi}. Independent Q-learning (IQL) \cite{tan1993multi} represents the most straightforward and widely utilized methodology in this context.  In IQL, each local Q-function is solely dependent on the local action, i.e., $Q_i(s, a) \approx Q_i(s, a_i)$. Similar to IQL, an alternative actor-critic version of MARL known as Independent Advantage Actor-Critic (IA2C) has been proposed in \cite{chu2019model}. While IQL and IA2C present fully scalable solutions, they encounter difficulties in dealing with partial observability and non-stationary MDP, primarily due to its inherent assumption that all other agents' behaviors form part of the environmental dynamics, even though their policies are continually updated during the training process \cite{chu2019multi}. 

To tackle the non-stationary and partial-observability issues prevalent in MARL, in \cite{zhang2018fully}, the critic network is fully decentralized but each agent takes global observations and actions and then performs consensus updates. Although their approach eliminates the need for a centralized controller during the training phase, it still necessitates access to global information. Several studies have focused on leveraging communication to address the issue of partial observability. For instance, FPrint \cite{foerster2017stabilising} investigates the impact of direct communication among agents, demonstrating that sharing low-dimensional policy fingerprints can enhance performance. In DIAL \cite{foerster2016learning}, each DQN agent generates the communication message together with action-value estimation, then the message is encoded and integrated with other input signals at the receiver's end. In contrast, CommNet \cite{sukhbaatar2016learning} offers a more generalized communication protocol, but it merely calculates the mean of all messages rather than encoding them. NeurComm \cite{chu2020multiagent, chen2021powernet} introduces a learnable communication protocol, where communication messages are encoded and concatenated to minimize information loss. However, these strategies generally implement a centralized controller (i.e., information aggregator or centralized critic networks) during the training and online adaptation phases, and the communication messages are typically raw or encoded network parameters, which often impose a burden on the communication channels due to the volume of information being transmitted.

To tackle the aforementioned issues, in this paper, we present a fully decentralized MARL algorithm for the CACC problems, which not only offers satisfactory control performance but also facilitates efficient communication through a quantization-based communication protocol. Section~\ref{sec:5} provides performance comparisons between the proposed algorithm and the previously mentioned benchmarks, demonstrating the effectiveness and potential benefits of our approach.

\section{Cooperative Adaptive Cruise Control (CACC)} \label{sec:3}
In this section, we introduce the system model for vehicle platooning along with the behavior model employed within the platoon. Furthermore, we present two representative CACC scenarios considered in this paper.

\begin{figure}[]
  \centering
  \includegraphics[width=0.48\textwidth]{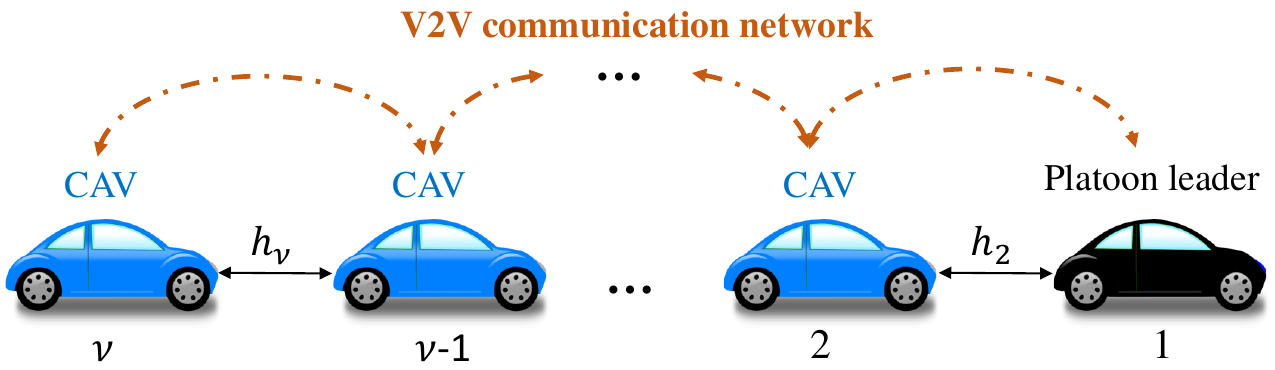}
  \caption{Framework of the CACC system.}
  \label{fig:platoon}
  \vspace{-10pt}
\end{figure}

\vspace{-10pt}
\subsection{Vehicle Dynamics}
As shown in Figure~\ref{fig:platoon}, we consider a platoon, comprising $\mathcal{V}$ CAVs, driving along a straight road. For simplicity, we assume that all vehicles in the system share identical characteristics such as maximum allowed acceleration and deceleration. The platooning system is guided by a platoon leader vehicle (PL, 1$st$ vehicle), while the platoon member vehicles (PMs, $i \in {2, ..., \mathcal{V}}$) travel behind the PL. Each PM $i$ maintains a desired inter-vehicle distance (IVD) $h_i$ and velocity $v_i$ relative to its preceding vehicle $i-1$, based on its unique spacing policy \cite{zhu2022joint}. The one-dimensional dynamics of vehicle $i$ can be expressed as follows:
\begin{subequations}
\begin{IEEEeqnarray}{cc}
\dot{h}_i = v_{i-1} - v_i,\\
\dot{v}_i = u_i,
\end{IEEEeqnarray}  
\end{subequations}
where $v_{i-1}$ and $u_i$ symbolize the velocity of its preceding vehicle and the acceleration of vehicle $i$, respectively. As per the design outlined in \cite{chu2019model}, the discretized vehicle dynamics, given a sampling time $\Delta t$, can be described by
\begin{subequations}
\begin{IEEEeqnarray}{cc}
h_{i, t+1} = h_{i, t} + \int_{t}^{t + \Delta t} (v_{i-1, \tau} - v_{i, \tau}) d \tau,\\
v_{i, t+1} = v_{i, t} + u_{i, t} \Delta t.
\end{IEEEeqnarray}  
\end{subequations}

In order to guarantee both comfort and safety, each vehicle must follow the following constraints \cite{chu2019model}:
\begin{subequations}
\begin{IEEEeqnarray}{ccc}
h_{i, t} \geq h_{min},\\
0 \leq v_{i,t} \leq v_{max},\\
u_{min} \leq u_{i,t} \leq u_{max},
\end{IEEEeqnarray}  
\end{subequations}
where $h_{min} = 1$ m, $v_{max}=30$ m/s, $u_{min} = - 2.5~m/s^2 < 0$ and $u_{max} = 2.5~m/s^2 > 0$ represent the minimum safe headway, maximum speed, deceleration, and acceleration limits, respectively.

\vspace{-10pt}
\subsection{Vehicle Behavior}\label{sec:ovm}
The behaviors of vehicles in the platoon are simulated using the optimal velocity model (OVM)~\cite{bando1995dynamical}. The OVM has been widely used in traffic flow modeling due to its ability to capture real human driving behaviors \cite{chu2020multiagent}. The principal equation of OVM for the $i$th vehicle is defined as follows:
\begin{equation} \label{eqn:ovm1}
u_{i, t} = \alpha_i (v^\circ (h_{i, t}; h^s, h^g) - v_{i, t}) + \beta_i (v_{i-1, t} - v_{i,t}),   
\end{equation}
where $\alpha_i$ and $\beta_i$ are the headway gain and relative velocity gain, respectively. These parameters serve as representations of human driver behavior, encapsulating the influence of both spacing and relative speed in determining vehicle acceleration. Here, $h^s=5$ m and $h^g=35$ m denote the stop headway and full-speed headway, both of which are key to understanding traffic dynamics at different vehicle densities. Furthermore, $v^\circ$ represents the headway-based velocity policy, which is defined as:
\begin{equation}  \label{eqn:ovm2}
     v^\circ (h) \triangleq 
\begin{cases} 
      0, & \text{if } h < h^s, \\
      \frac{1}{2} v_{max} (1 - \cos{(\pi \frac{h - h^s}{h^g - h^s})}), & \text{if } h^s \leq h \leq h^g,\\
      v_{max}, & \text{if } h > h^g . \\
   \end{cases}
\end{equation}
This policy function serves as an optimal velocity strategy for each vehicle based on the current headway to the preceding vehicle. At small headways less than or equal to $h^s$, the optimal velocity is zero, highlighting the need for the ego vehicle to stop for preventing potential collisions. For headways within the range $h^s$ to $h^g$, the optimal velocity gradually increases by following a cosine curve until reaching the maximum velocity. For large headways greater than or equal to $h^g$, the optimal velocity is capped at the vehicle's maximum speed, ensuring both safety and efficiency in the traffic flow. This strategy significantly contributes to maintaining fluidity in vehicular traffic under various density conditions. The OVM will be used to simulate the behavior of vehicles where the RL will train the driving hyper-parameters $(\alpha_i,\beta_i)$. See Section~IV for more details on the RL formulation. In this paper, our primary focus is on validating the fully decentralized MARL approach in the context of CACC. As for the exploration of more complex vehicle dynamics \cite{feng2019string}, this will be a key area of investigation in our future work.

\vspace{-10pt}
\subsection{Two CACC Scenarios}\label{sec:2cacc}
In this paper, following \cite{chu2020multiagent} two different CACC scenarios are investigated: ``Catchup'' and ``Slowdown''. Recall that the objective of CACC is to adaptively control a fleet of CAVs in order to reduce the car-following headway to a pre-specified value (e.g., $h^*=20$ m) and achieve a target velocity (e.g., $v^*=15$ m/s), by leveraging real-time V2V communications.   For the ``Catchup'' scenario, the platoon members (PMs) ($i = 2, ..., \mathcal{V}$) are initialized with states $v_{i, 0} = v^*_t$ and $h_{i, 0} = h^*_t$, while the platoon leader (PL) is initialized with states $v_{1, 0}= v^*_t$ and $h_{1, 0} = a \cdot h^*_t$, where $a$ is a random variable uniformly distributed between 1.5 and 2.5. In contrast, during the ``Slowdown'' scenario, all vehicles ($i = 1, ..., \mathcal{V}$) have initial velocities $v_{i, 0} = b \cdot v^*_t$ and $h_{i, 0} = h^*_t$, where $b$ is uniformly distributed between 1.5 and 2.5. Here, $v^*_t$ linearly decreases to 15 m/s within the first 30 seconds and then remains constant. The ``Slowdown'' scenario poses a more complex and challenging task than the ``Catchup'' scenario due to the necessity for all vehicles to coordinate their deceleration rates and maintain safe inter-vehicle distances with higher accuracy, thereby requiring more precise control strategies. Examples of the headway and speed profiles of the CAVs in these scenarios are illustrated in Figures~\ref{fig:profiles_Catchup} and \ref{fig:profiles_Slowdown}.

\section{CACC as MARL}\label{sec:4}\
In this section, we first formulate the considered CACC problem as a partially observable Markov decision process (POMDP). Subsequently, we present our fully decentralized MARL algorithm, which represents our primary strategy for addressing the challenges presented in the CACC problem. Then, we introduce the quantization-based communication protocol to enhance the efficiency of agent communication in the MARL framework without major performance degradation.

\vspace{-7pt}
\subsection{MARL Formulation}
In this paper, we model the CACC problem as a model-free multi-agent network \cite{chu2020multiagent}, where each agent (i.e., AV)  is capable of communicating with the vehicles ahead and behind via V2V communication channels. We denote the global state space and action space as $S := \times_{i \in \text{\larger[2]$\nu$}} S_i$ and $\mathcal{A} := \times_{i \in \text{\larger[2]$\nu$}} \mathcal{A}_i$, respectively. The intrinsic dynamics of the system can be characterized by the state transition distribution $\mathcal{P}$: $\mathcal{S} \times \mathcal{A} \times \mathcal{S} \rightarrow [0, 1]$. We propose a fully decentralized MARL framework where each agent $i$ (equivalently, AV $i$) has a partial view of the environment, specifically the surrounding vehicles, which accurately reflects the practical scenario where AVs are limited to sensing or communicating with neighboring vehicles, thereby rendering the overall dynamical system as a POMDP. This POMDP, $\mathcal{M_G}$, can be delineated by the following tuple $\mathcal{M_G} = (\{\mathcal{A}_i, \mathcal{S}_i, \mathcal{R}_i\}_{i\subseteq \mathcal{V}+1}, \mathcal{T})$:

\begin{itemize}
\item \textbf{Action space}: In the considered CACC problem, the action $a_t \in \mathcal{A}_i$ is straightforwardly related to the longitudinal control. However, due to the data-driven nature of RL, formulating a safe and robust longitudinal control strategy poses a significant challenge \cite{chu2019model}. To address this, we adopt OVM (see Section~\ref{sec:ovm}, \cite{bando1995dynamical}) to carry out the longitudinal vehicle control. The OVM control behavior is affected by various hyperparameters: headway gains $\alpha$, relative velocity gain $\beta$, stop headway $h^s$, and full-speed headway $h^g$. Usually, ($\alpha$; $\beta$) represents the driving behavior of a human driver. However, following \cite{chu2020multiagent}, we leverage MARL to propose suitable values of ($\alpha$; $\beta$) for each OVM controller. These recommended values are selected from a set of four different levels: \{(0, 0), (0.5, 0), (0, 0.5), (0.5, 0.5)\}. Subsequently, the longitudinal action can be computed using Eq.~\ref{eqn:ovm1} and Eq.~\ref{eqn:ovm2}.

\item \textbf{State space}: The state space represents the description of the environment. The state of agent $i$, $\mathcal{S}_i$, is defined as $[v, v_{diff}, vh, h, u]$, where $v=(v_{i,t} - v_{i,0}) / v_{i,0}$ denotes the current normalized vehicle speed. $v_{diff} = \text{clip}((v_{{i-1},t} - v_{i,t}) / 5, -2, 2)$ represents clipped vehicle speed difference with its leading vehicle. $vh = \text{clip} ((v^\circ (h) - v_{i,t}) / 5, -2, 2)$, $h = (h_{i,t} + (v_{i-1, t} - v_{i,t}) \Delta t - h^*) / h^*$, and $u = u_{i, t} / u_{max}$ are the headway-based velocity defined in Eq.~\ref{eqn:ovm2}, normalized headway distance, and acceleration, respectively.

\item \textbf{Reward function}: The reward function $r_{i,t}$ is pivotal for training the RL agents to exhibit the desired behaviors. With our objective being the training of our agents to achieve a predefined car-following headway $h^*=20$ m and velocity $v^*=15$ m/s, the reward assigned to the $i$th agent at each time step $t$ is designed as follows:
\begin{equation} \label{eqn:reward_fn}
\begin{aligned}
r_{i, t} &= w_1 (h_{i, t} - h^*)^2 + w_2 (v_{i, t} - v^*)^2 
\\
&\quad +  w_3 u^2_{i, t} + w_4 (2h_{s} - h_{i, t})^2_{+}, \end{aligned}  
\end{equation}
where $w_1$, $w_2$, $w_3$, and $w_4$ are the weighting coefficients. In this equation, the first two terms, $(h_{i, t} - h^*)^2$ and $(v_{i, t} - v^*)^2$, penalize deviations from the desired headway and velocity, encouraging the agent to achieve these targets closely. The third term, $u^2_{i, t}$, is included to minimize abrupt accelerations, thereby promoting smoother and more comfortable rides for passengers. Lastly, the term $(2h_{s} - h_{i, t})^2_{+}$ functions as a safety constraint, penalizing the agent heavily if the inter-vehicle distance is less than twice the stop headway $h_{s}$, which is critical for preventing collisions and ensuring the safety of the vehicle platoon. The ``+'' operator represents that this term contributes to the reward only when $h_{i, t}$ is less than $2h_{s}$, similar to the Rectified Linear Unit (ReLU) function. This comprehensive reward design serves to balance performance, comfort, and safety considerations in the CACC system. Upon a collision, if the inter-vehicle distance $h_{i, t} \leq 1$ m, each agent is subjected to a substantial penalty of 1000, resulting in immediately terminating the training episode.

\item \textbf{Transition probabilities}: The transition probability, $\mathcal{T}(s'|s,a)$, characterizes the underlying dynamics of the system. Given that our approach is a model-free MARL framework, we do not assume any prior knowledge of this transition probability while developing our MARL algorithm.
\end{itemize}

\vspace{-10pt}
\subsection{Fully Decentralized MARL}
In this paper, we formulate the CACC as a \textit{fully decentralized} MARL problem, where each agent (i.e., an AV) independently decides its action based on its local observation during both training and execution. Importantly, this structure does not require a centralized controller, meaning that each agent possesses its own individual policy networks. During the learning phase, agents rely on locally received rewards to train and update these networks. In this paper, we employ a MARL framework in which each agent is equipped with its actor-critic network \cite{mnih2016asynchronous}, and the policy network for agent $i$ is updated with gradient ascend and the gradient is defined as:
\begin{equation}\label{eqn:advantagepolicygradient_MARL}
\nabla_{\theta} \mathcal{L}(\pi_{\theta_i}) = E_{\pi_{\theta_i}} \left[\sum_{t=0}^T \nabla_{\theta} \log \pi_{\theta_i}(a_{i, t}|s_{i, t})
 A^{\pi_{\theta_i}}_{i, t} \right],
\end{equation}
where $A^{\pi_{\theta_i}}_{i, t} = r_{i,t} + \gamma V^{\pi_{\phi_i}}(s_{i, t+1}) - V^{\pi_{\phi_i}}(s_{i, t})$ is the advantage function  and $V^{\pi_{\phi_i}}(s_{i,t})$ is the state value function, which is updated following the loss function:
\begin{equation}\label{eqn:valueloss_MARL}
\mathcal{L}^{V^{\pi_{\phi_i}}} = \min_{\phi_i} E_{\mathcal{D}_i} \Big [r_{i, t} + \gamma V^{\pi_{\phi_i}}(s_{i, t+1}) - V^{\pi_{\phi_i}}(s_{i, t})\Big ]^2.
\vspace{-5pt}
\end{equation}

\begin{figure}[!ht]
  \centering
  \includegraphics[width=0.42\textwidth]{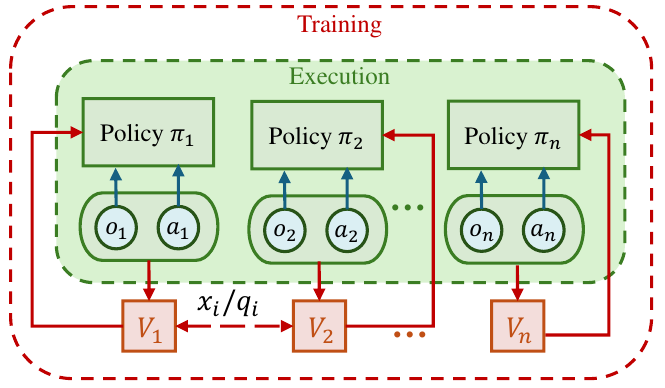}
  \caption{Overview of our fully-decentralized MARL approach.}
  \label{fig:MACACC}
  \vspace{-10pt}
\end{figure}

Despite each agent learning independently, the overall goal of the cooperative MARL framework is to optimize the average global reward $r_{g,t} = \frac{1}{\mathcal{V}} \sum_{i=1}^{\mathcal{V}} r_{i, t}$. To address the non-stationary, in \cite{zhang2018fully}, the update of the policy network is executed independently by each agent, eliminating the need for inferring other agents' policies. However, when it comes to updating the critic network, a collaborative approach is adopted, in which each agent shares its estimate of the value function $x_{i}$ with its neighboring agents within the network through a ``mean'' operation, i.e., $x_{i}^{k+1} = \frac{1}{|\mathcal{N}_i|} \sum_{j \in \mathcal{N}_i} x_{j}^k$, where $\mathcal{N}_i$ is the neighboring set of agent $i$. This allows for the joint evolution and continuous improvement of the system's overall performance. However, their approach is based on the assumption that all agents are homogeneous, sharing the same characteristics. While this simplifies the problem structure, it does not adequately represent the intrinsic diversity of individual agents, which is particularly relevant for the CACC scenario where diverse strategies are needed based on vehicles' positions, speeds, and proximities. To address this concern, we propose a novel update strategy that fosters a balance between individual learning and collaborative influence from neighboring agents. Specifically, we assume that the agents interact on an undirected graph, and the interaction can be described by a weight matrix $W$. If agent $i$ and agent $j$ can communicate and interact with each other, then the $(i, j)$-th entry of $W$, i.e., $w_{ij}$, is positive (e.g., 1.0). Otherwise, $w_{ij}$ is zero. During the training, the update strategy is designed as
\begin{equation}\label{eqn:update1}
x_{i}^{k+1} = x_i^k + \epsilon \sum_{j \in \mathcal{N}_i} \omega_{ij} (x_{j}^k - x_{i}^k)-  \lambda g_i^k,
\end{equation}
where $g_{i}^{k}$ is the gradient that agent $i$ obtains at iteration $k$ for optimization, $\epsilon$ is the scaling factor used to modulate the impact or collaborative influence from neighboring agents, and $\lambda = 5.0 \times 10^{-4}$ is the learning rate that adjusts the influence of the gradient on the update process. This novel update strategy fosters collaboration among the agents while preserving the individual learning capabilities of each, thereby striking a balance between global performance optimization and localized adaptivity. An overview of our fully-decentralized MARL framework is given in Figure~\ref{fig:MACACC}.

\begin{remark}
The scaling factor $\epsilon$ plays an important role in the update scheme. If $\epsilon$ is too large, the agent allocates more influence from the neighboring agents. On the other hand, if $\epsilon$ is too small, the agent will concentrate more on its own updates. In this work, we use cross-validations and find that $\epsilon = 1.0 \times 10^{-3}$ and $\epsilon = 1.0 \times 10^{-4}$ are the optimal choices for ``Catchup'' and ``Slowdown'' scenarios, respectively.
\end{remark}

The update strategy of the proposed fully decentralized MARL for CACC (abbreviated as \textbf{MACACC}) is given in Algorithm~\ref{algo:marl1}. 

\begin{figure}[!ht]
\vspace{-10pt}
\removelatexerror
\scalebox{0.85}{
\begin{algorithm*}[H]
\SetAlFnt{\small}
    \SetKwInOut{Parameter}{Parameter}
    \SetKwInOut{Output}{Output}
\caption{MACACC for CACC}
\label{algo:marl1}
\SetAlgoLined
Public parameters: $W$, $\epsilon$, $\lambda$, $x_i^0$ for all $i$, the total number of iterations $k$

\For{$i$th agent}{
Determine the local gradient $g_i^k$ for the critic network;\\
Send critic estimate to all neighboring agents $j \in \mathcal{N}_i$;\\
After receiving $x_{j}^k$ from all $j \in \mathcal{N}_i$, update network parameters as
$$x_{i}^{k+1} = x_i^k + \epsilon \sum_{j \in \mathcal{N}_i} \omega_{ij} (x_{j}^k - x_{i}^k)- \lambda g_i^k$$} 
\end{algorithm*}}
\vspace{-10pt}
\end{figure}

\vspace{-10pt}
\subsection{Quantization-based Communication Protocol}
To enhance the communication efficiency among agents in our MARL framework, we propose a strategy of transmitting quantized parameters, rather than the raw parameters of the critic network. This approach is especially important for autonomous driving applications that are often subject to limited communication bandwidth. By transmitting compact, quantized parameters instead of raw data, we ensure optimal use of available bandwidth, thereby fostering efficient and effective communication among the vehicles in the network. Quantization-based techniques have been studied in distributed optimization and learning~\cite{alistarh2017qsgd,wang2022quantization}, however, to the best of our knowledge, this is the first work that incorporates a quantization-based communication protocol into the MARL framework for achieving communication-efficient CACC.

\begin{figure}[!ht]
  \centering
  \includegraphics[width=0.40\textwidth]{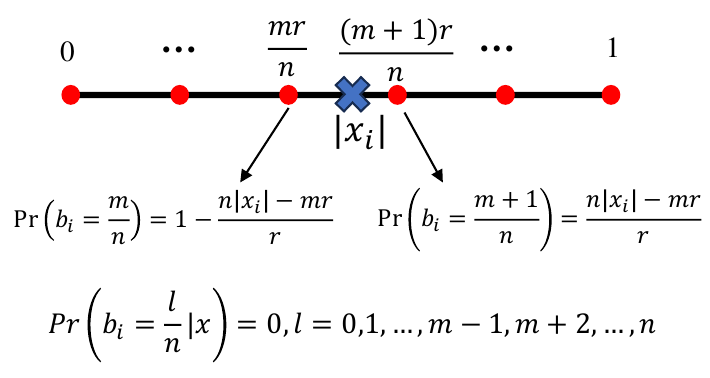}
  \caption{Illustration of the quantization-based communication protocol.}
  \label{fig:quantization}
  \vspace{-10pt}
\end{figure}

Let the parameters of the critic/value network be denoted as $x= [x_1, x_2, ..., x_d]^T$, with $d$ representing the dimension of the parameter vector. We then apply a quantization function $\mathcal{Q}(x)$ to these parameters, yielding a quantized parameter vector $[q_1, q_2, ..., q_d]^T$. The quantization rule (also see Figure~\ref{fig:quantization}) is defined as:
\begin{equation}\label{eqn:quan1}
q_i = r\cdot \text{sign}(x_i) b_i.
\end{equation}
In \eqref{eqn:quan1}, $r$ is a non-negative real number no less than the $\ell_{\infty}$ norm of $x$, and $\text{sign}(\cdot)$ represents the sign function, which returns the sign of any given real number. The factor $b_i$ is a random variable following a designed distribution determined by the magnitude of the corresponding parameter $x_i$. Let $n$ be the resolution of the quantization, and then $b_{i}$ is selected from the set $\{0, \frac{1}{n}, \frac{2}{n}, \cdots, 1\}$, indicating that $q_{i} \in \{-r, -\frac{n-1}{n}r, \cdots, 0, \cdots, \frac{n-1}{n}r, r\}$. Let $0\le m \le n-1$ be an integer such that $|x_i|$ belongs to the interval $[\frac{m}{n} r, \frac{m+1}{n} r]$. Then the probability distribution of $b_i$ is determined by
\begin{subequations} \label{eqn:quan2}
\begin{IEEEeqnarray}{ccc}
P(b_i=\frac{m+1}{n}|x) = \frac{n|x_i|-mr}{r}, \\
P(b_i=\frac{m}{n}|x) = 1 - \frac{n|x_i|-mr}{r}, \\
P(b_i=\frac{l}{n}|x) = 0, l=0, \cdots, m-1, m+2, \cdots, n.
\end{IEEEeqnarray}  
\end{subequations}
It can be concluded that if the magnitude of $|x_i|$ is closer to $\frac{m+1}{n} r$, then the higher the probability that $b_i$ will be $\frac{m+1}{n}$, and vice versa. We denote the quantization-based MACACC algorithm as \textbf{QMACACC ($n$)}. An extremely condensed version of QMACACC is QMACACC (1), in which only three discrete numbers $\{-r, 0, r\}$ are used to represent each parameter, and $b_i$ is defined as:
\begin{subequations} \label{eqn:quan3}
\begin{IEEEeqnarray}{ccc}
P(b_i=1|x) = \frac{|x_i|}{r}, \\
P(b_i=0|x) = 1-\frac{|x_i|}{r}.
\end{IEEEeqnarray}  
\end{subequations}

\begin{remark}
The quantization resolution $n$ is a crucial hyperparameter in the quantization scheme. When $n$ is small, the sparse quantization intervals could lead to excessive loss of information, negatively impacting the performance of the MARL framework. Conversely, when $n$ is large, the computational and communication overheads could increase due to the larger number of potential quantized values. Therefore, choosing an appropriate value of $n$ is crucial for balancing communication efficiency and the performance of the MARL framework. An empirical evaluation of different $n$ values will be conducted in Section~\ref{sec:5}.
\end{remark}

The update strategy of the proposed quantization-based MARL (i.e., QMACACC ($n$)) for CACC is given in Algorithm~\ref{algo:marl2}. 

\begin{figure}[!ht]
\vspace{-10pt}
\removelatexerror
\scalebox{0.85}{
\begin{algorithm*}[H]
\SetAlFnt{\small}
    \SetKwInOut{Parameter}{Parameter}
    \SetKwInOut{Output}{Output}
\caption{QMACACC ($n$) for CACC}
\label{algo:marl2}
\SetAlgoLined
Public parameters: $W$, $\epsilon$, $\lambda$, $x_i^0$ for all $i$, the total number of iterations $k$

\For{$i$th agent}{
Determine the local gradient $g_i^k$ for the critic network;\\
Quantize  critic estimate according to Eqs.~\ref{eqn:quan1} and ~\ref{eqn:quan2} and send to all neighboring agents $j \in \mathcal{N}_i$;\\
After receiving $\mathcal{Q}(x_{j}^k)$ from all $j \in \mathcal{N}_i$, update state as
$$x_{i}^{k+1} = x_i^k + \epsilon \sum_{j \in \mathcal{N}_i} \omega_{ij} (\mathcal{Q}(x_{j}^k) - \mathcal{Q}(x_{i}^k))- \lambda g_i^k$$} 
\end{algorithm*}}
\vspace{-10pt}
\end{figure}

\begin{figure*}[]
  \centering
  \includegraphics[width=0.8\textwidth]{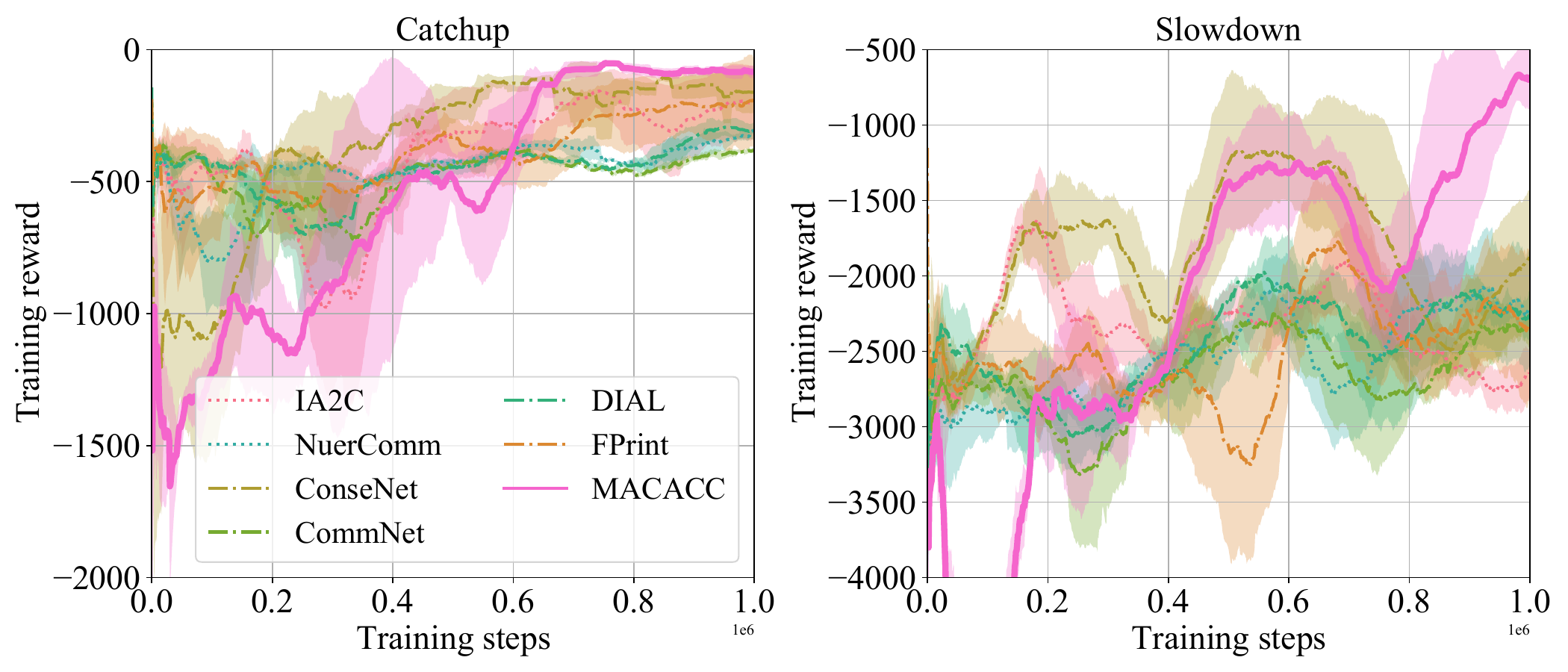}
  \caption{Training curves comparison between the proposed MARL policy (MACACC) and 6 state-of-the-art MARL benchmarks.}
  \label{fig:train_reward}
  \vspace{-10pt}
\end{figure*}

\begin{table*}[]
\renewcommand{\arraystretch}{1.4}
\centering
\caption{Average execution performance comparison over trained MARL policies. The best values are in bold.}
\label{tab:benchmarks_multi}
\begin{tabular}{cccccccc}
\hline
\textbf{Scenario Name} & \textbf{IA2C} & \textbf{FPrint} & \textbf{ConseNet} & \textbf{NeurComm} & \textbf{CommNet} & \textbf{DIAL} & \textbf{MACACC} \\ \hline \hline
Catch-up               & -241.38  & -198.93            & -94.67              & -301.41     & -397.55             & -227.68        & \textbf{-50.44}       \\ 
Slow-down              & -2103.38  & -1470.41            & -1746.43              &      -1912.23        & -2590.72            & -1933.27          & \textbf{-492.30}            \\ \hline 
\end{tabular}
\end{table*}

\begin{table*}[]
\renewcommand{\arraystretch}{1.4}
\centering
\caption{Performance of MARL controllers in CACC environments: ``Catchup'' (above) and ``Slowdown''
(below). The best values are in bold.}
\label{tab:benchmarks_metrics}
\begin{tabular}{lccccccc}
\hline 
\textbf{Temporal Average Metrics} & \textbf{IA2C} & \textbf{FPrint} & \textbf{ConseNet} & \textbf{NeurComm} & \textbf{CommNet} & \textbf{DIAL} & \textbf{MACACC} \\ \hline \hline
avg vehicle headway {[}m{]}       & 19.43         & \textbf{20.02}             & 20.28                & 21.77                & 22.38              & 21.86            & 19.91              \\ 
avg vehicle velocity {[}m/s{]}    & \textbf{15.00}            & 15.34              & 15.30               & 15.04                & 15.01               & 15.01            & 15.32              \\ 
collision number                  & 0             & 0              & 0                & 0               & 0               & 0            & \textbf{0}                  \\ \hline
avg vehicle headway {[}m{]}       & -             & 15.16              & 9.23               & 11.45                & 4.90               & 9.71            & \textbf{20.44}              \\ 
avg vehicle velocity {[}m/s{]}    & -             & 13.10              & 8.08                & 10.32                & 4.18              & 8.91            & \textbf{16.61}              \\ 
collision number                  & 50            & 14              & 29                & 22             & 38               &    26         & \textbf{0}                  \\ \hline
\end{tabular}
\vspace{-10pt}
\end{table*}

\begin{figure}
\centering
\subfloat[Headway profiles.]{\label{4figs-a} \includegraphics[width=0.235\textwidth]{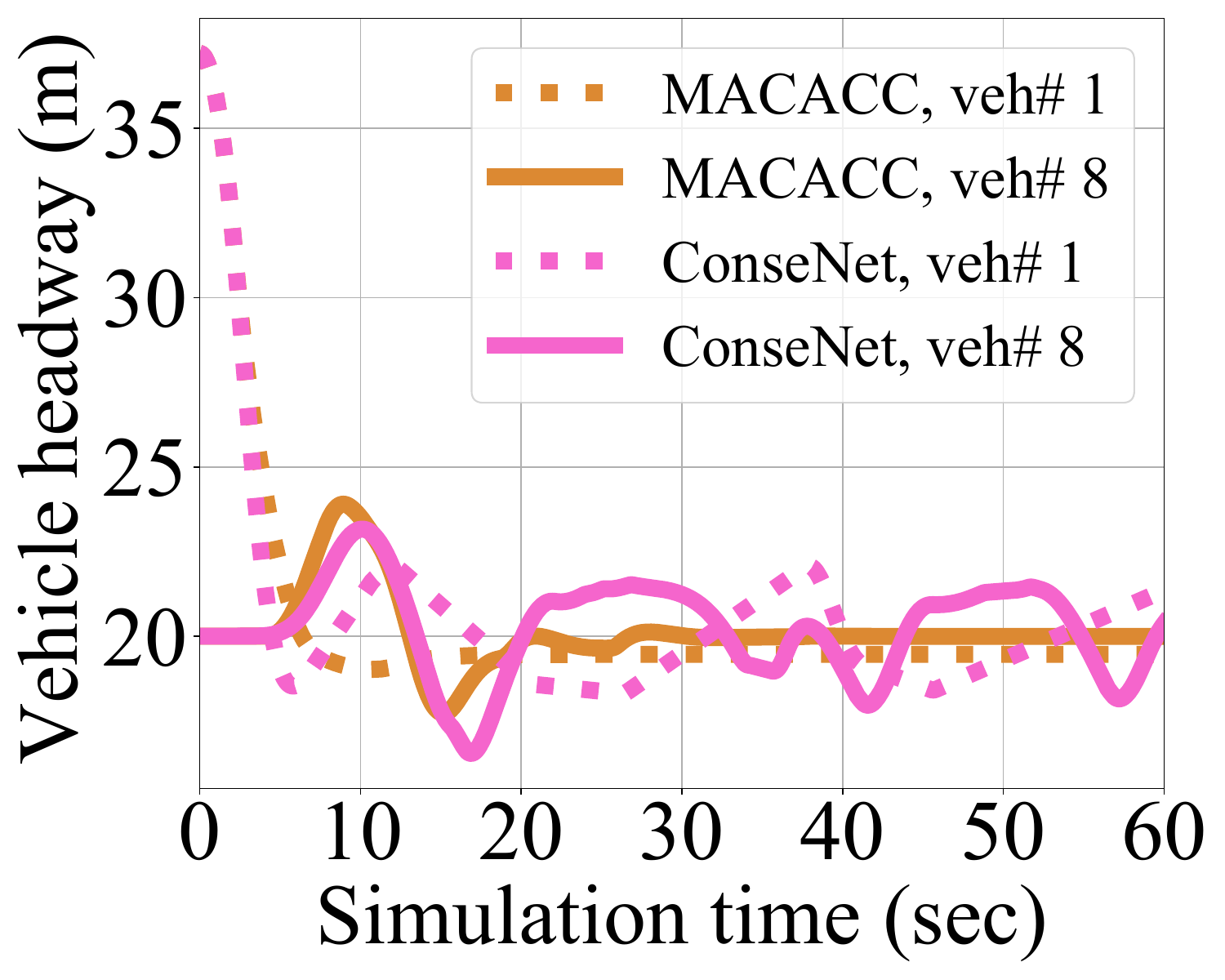}}%
\hfill
\subfloat[Velocity profiles.]{\label{4figs-b} \includegraphics[width=0.235\textwidth]{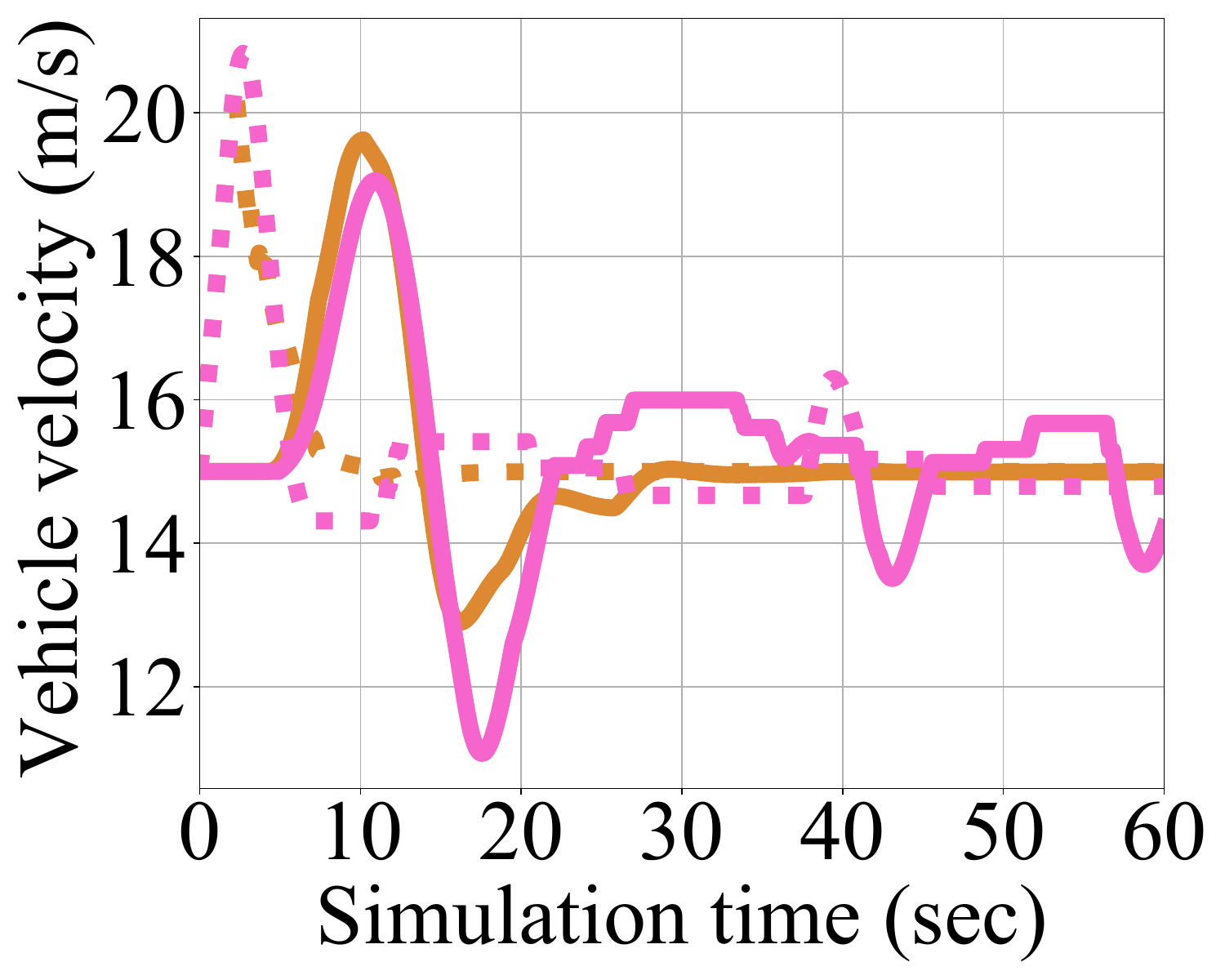}}%
\caption{Headway and velocity profiles in the ``Catchup'' environment of the first and last vehicles of the platoon, controlled by the proposed approach (MACACC) and the top baseline policy (ConseNet).}
\label{fig:profiles_Catchup}
\vspace{-10pt}
\end{figure}

\begin{figure}
\centering
\subfloat[Headway profiles.]{\label{4figs-a2} \includegraphics[width=0.235\textwidth]{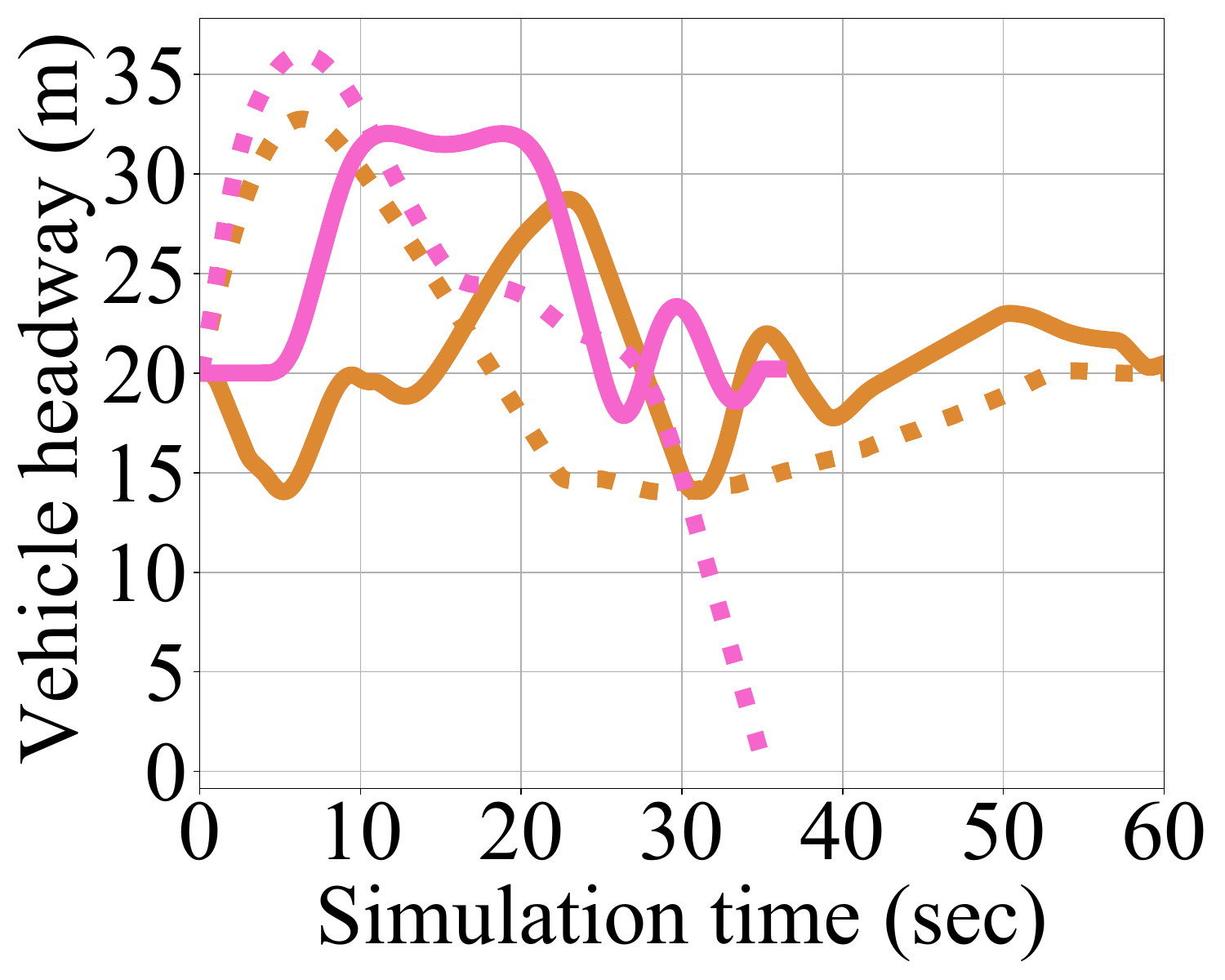}}%
\hfill
\subfloat[Velocity profiles.]{\label{4figs-b2} \includegraphics[width=0.235\textwidth]{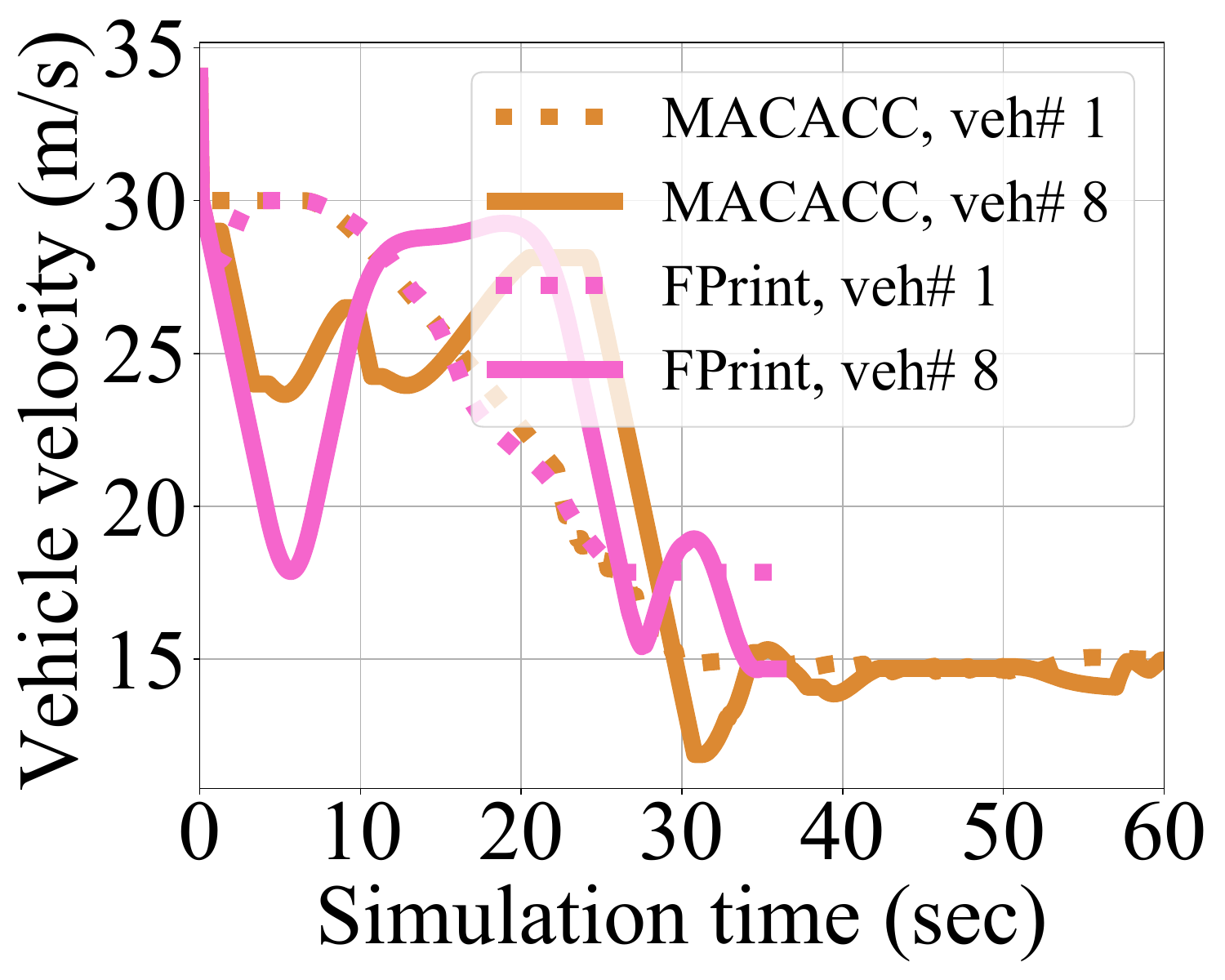}}%
\caption{Headway and velocity profiles in the ``Slowdown'' environment of the first and last vehicles of the platoon, controlled by the proposed approach (MACACC) and the top baseline policy (FPrint).}
\label{fig:profiles_Slowdown}
\vspace{-10pt}
\end{figure}

\begin{figure*}[]
  \centering
  \includegraphics[width=0.8\textwidth]{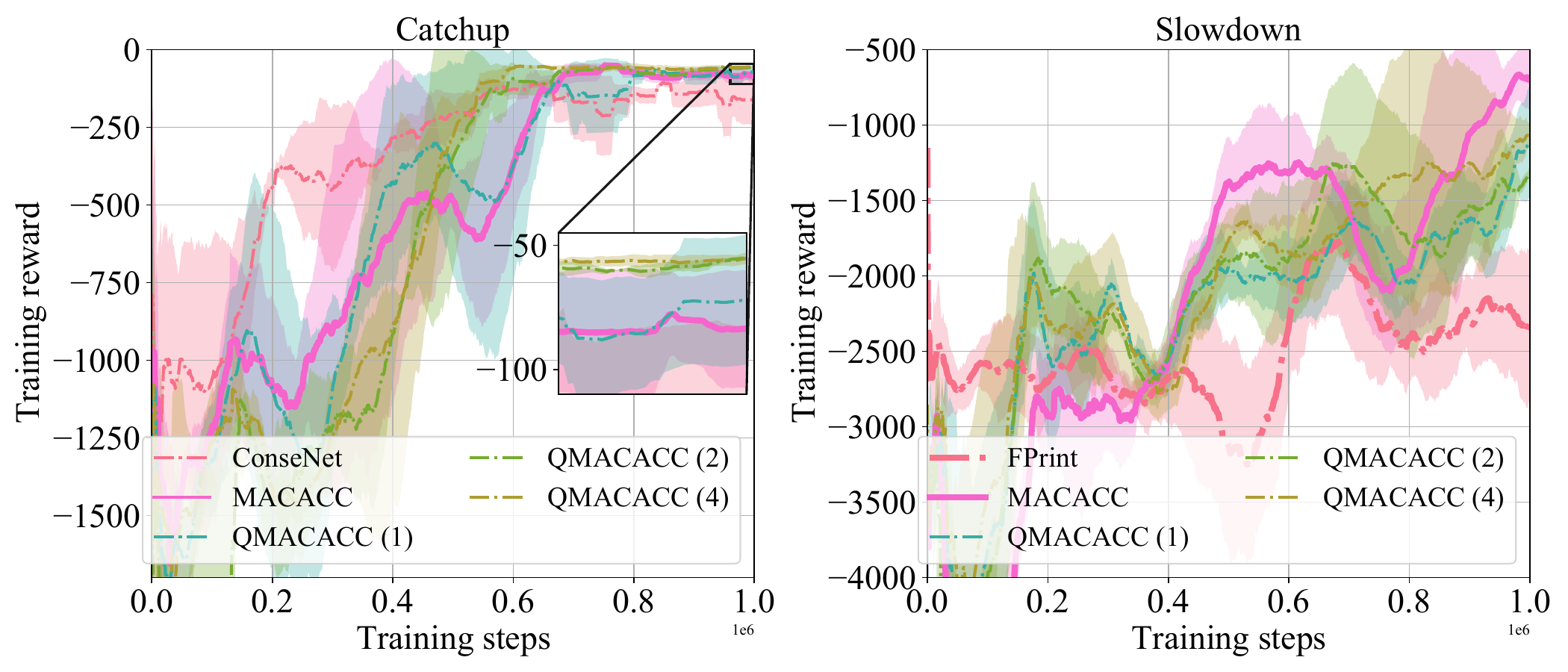}
  \caption{Training curves comparison between the proposed MARL policy (MACACC) and Quantization-based MACACC (QMACACC ($n$)).}
  \label{fig:train_reward_q}
  \vspace{-10pt}
\end{figure*}


\begin{figure*}[]
  \centering
  \includegraphics[width=0.8\textwidth]{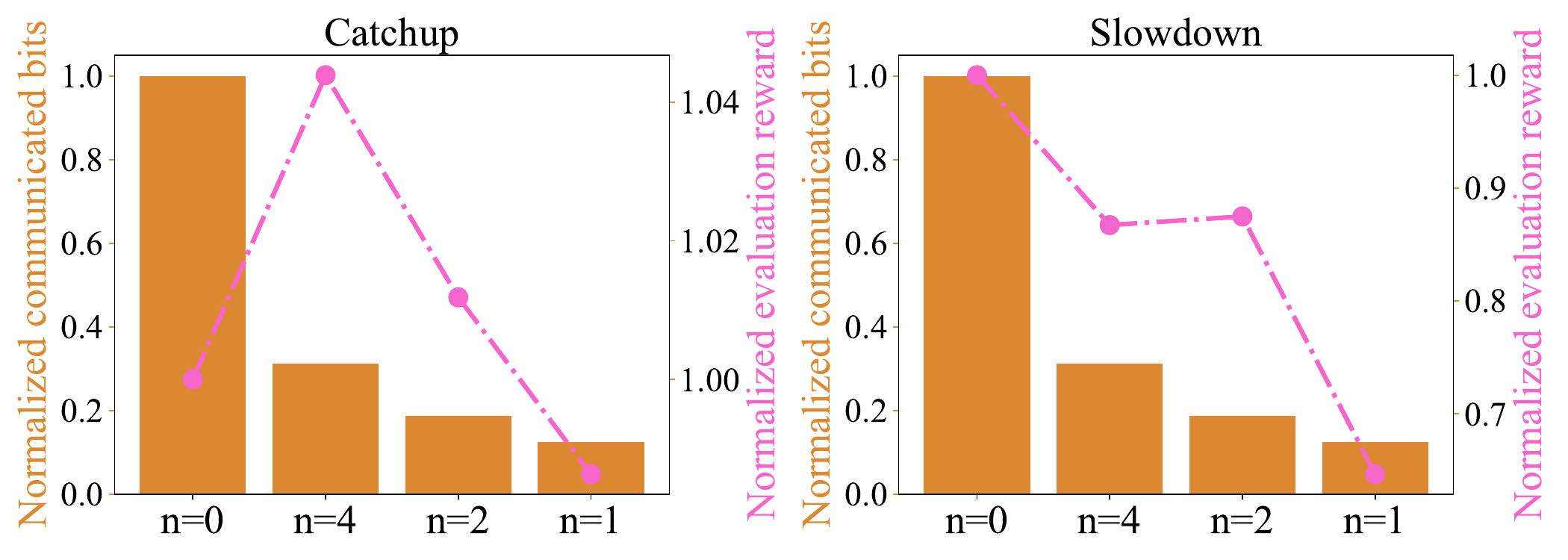}
  \caption{Transmitted message and execution performance comparison between the proposed MARL policy (MACACC ($n=0$)) and Quantization-based MACACC (QMACACC ($n=1, 2, 4$)).}
  \label{fig:comm_efficient}
  \vspace{-10pt}
\end{figure*}

\begin{figure*}[!ht]
  \centering
  \includegraphics[width=0.9\textwidth]{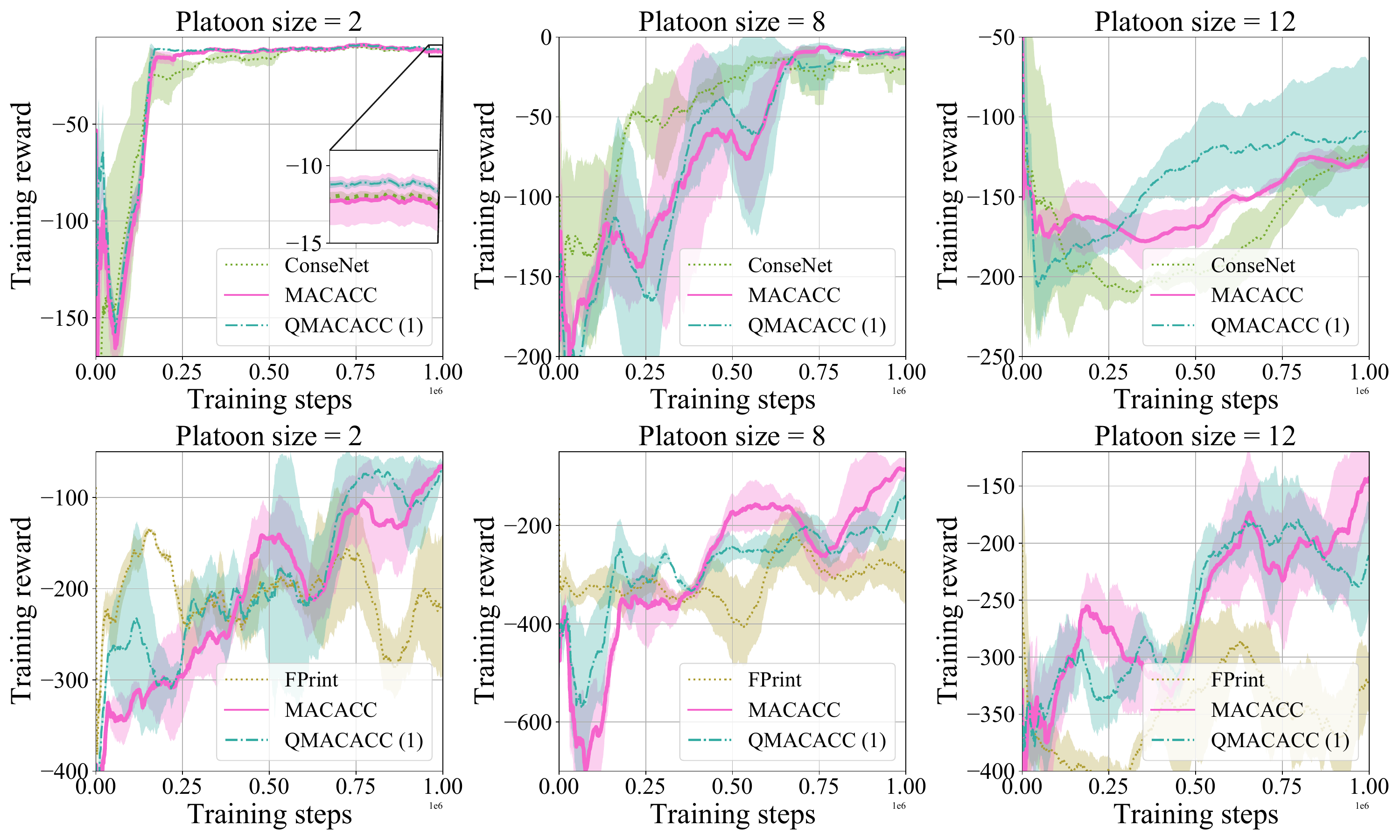}
  \caption{Normalized training curves comparison between MACACC, QMACACC (1), and the top baseline methods with different platoon sizes in the CACC scenarios: ``Catchup'' (above) and ``Slowdown'' (below).}
  \label{fig:train_reward_psize}
  \vspace{-10pt}
\end{figure*}

\section{Experimental Results \& Discussions}\label{sec:5}
In this section, we evaluate our MARL framework in two CACC scenarios detailed in Section~\ref{sec:2cacc}. Firstly, we benchmark our approach against several state-of-the-art MARL strategies. Then, we demonstrate the effectiveness of our quantization-based communication protocol.

\vspace{-10pt}
\subsection{General Setups}
To demonstrate the efficiency and robustness of the proposed approach, we compare it with several state-of-the-art MARL benchmark controllers discussed in Section~\ref{sec:2marl}. Specifically, IA2C performs independent learning, while ConseNet \cite{zhang2018fully} takes the ``mean" operation during updating critic networks, and FPrint \cite{foerster2017stabilising} incorporates the neighbors' policy into the inputs. DIAL \cite{foerster2016learning}, CommNet \cite{sukhbaatar2016learning} and NeurComm \cite{chu2020multiagent}, on the other hand, are implementations with learnable communication protocols, incorporating more messages from the neighbors, e.g., neighboring states or policy information, relying on higher communication bandwidth. All algorithms use the same DNN structures: one fully-connected layer for input state encoding and one LSTM layer for message extracting. All hidden layers have 64 units. During the training, the network is initialized with the state-of-the-art orthogonal initializer \cite{saxe2013exact}. We train each model over 1M steps, with $\gamma = 0.99$, actor learning rate $5.0 \times 10^{-4}$, and critic learning rate $2.5 \times 10^{-4}$. Also, each algorithm is trained three times with different random seeds for generalization purposes. Each training takes about 12 hours on a Ubuntu 18.04 server with an AMD 9820X processor and 64 GB memory. 

The hyperparameter $w_1$, $w_{2}$, $w_{3}$, and $w_{4}$ in the reward function~\eqref{eqn:reward_fn} are set to -1.0, -1.0, -0.1, and -5.0, respectively, with a significant emphasis on penalizing situations where the safety headway distance is insufficient. Considering a simulated traffic environment over a period of $T=60$ seconds, we define $\Delta t=0.1$ seconds as the interaction period between RL agents and the traffic environment, so that the environment is simulated for $\Delta t$ seconds after each MDP step. In the following experiments, we assume the platoon size to be $\mathcal{V} = 8$, implying that there are a total of 8 CAVs in the platoon. The impact of different platoon sizes on our model's performance will be studied and presented in Section~\ref{sec:5-4}.

\vspace{-10pt}
\subsection{Comparison with State-of-the-Art Benchmarks}
Figure~\ref{fig:train_reward} shows the performance comparison in terms of the learning curves between the proposed approach MACACC and several state-of-the-art MARL benchmarks. As expected, the proposed approach achieves the best performance, evidenced by higher training rewards in both CACC scenarios. In the more challenging ``Slowdown'' environment, the proposed approach shows its greater advantages of sample efficiency as seen from the fastest convergence speed and best training reward compared to other algorithms.

After training, we evaluate each algorithm 50 times with different initial conditions. Table~\ref{tab:benchmarks_multi} shows the evaluation performance comparison over the trained MARL policies. The proposed method consistently outperforms the benchmarks in all CACC scenarios in terms of the evaluation reward, which reveals the overall evaluation metrics including vehicle headway, velocity, acceleration, and safety as described in Eq.~\ref{eqn:reward_fn}. 
Table~\ref{tab:benchmarks_metrics} shows the key evaluation metrics in CACC. The best headway and velocity averages are the closest ones to $h^* = 20$ m, and $v^* = 15$ m/s. Note the averages are only computed from safe execution episodes, and we use another metric ``collision number'' to count the number of episodes where a collision happens within the horizon. Ideally, ``collision-free'' is the top priority. It is clear that our approach achieves promising performance in the ``Catchup'' environment, and the best performance in the harder ``Slowdown'' environment. All algorithms achieve relatively good performance in the ``Catchup'' environment with zero collision number. It is surprising that IA2C achieves excellent average vehicle velocity at $v^*$. However, it demonstrates high collision numbers (i.e., 50) in the ``Slowdown'' scenario due to non-stationary issues since there is no communication between agents. FPrint yields the best average vehicle headway in the ``Catchup'' environment, while it has 14 out of 50 collisions during the testing. On the other hand, NeurComm and CommNet show great average vehicle velocity in the ``Catchup'' environment, however, they failed to track the optimal headway, resulting in average high headway of 21.77 m and 22.38 m, respectively. It is noted that ConseNet achieves promising performance in the ``Catchup'' environment, with a zero collision rate, average vehicle headway (20.28 m), and velocity (15.30 m/s) close to the optimal values. However, it yields high collision numbers (29 out of 50) in the ``Slowdown'' scenario as it simply encourages all agents to behave similarly via the ``average'' operations during training, which is especially impractical for complex scenarios, such as ``Slowdown'', where agents need to react differently to the speed and headway changes.

Figures~\ref{fig:profiles_Catchup} and \ref{fig:profiles_Slowdown} show the corresponding headway and velocity profiles for the selected controllers for the two CACC scenarios. In the ``Catchup'' scenario, as expected, the MACACC controller is able to achieve steady state $v^*$ and $h^*$ for the first and last vehicles of the platoon, whereas the ConseNet controller still faces difficulties in eliminating the perturbation through the platoon. In a harder ``Slowdown'' environment, MACACC is still able to achieve optimal headway at about 60 seconds and reach the optimal velocity quickly. However, FPrint fails the control task with a collision that happened at about 35 seconds. This may be because simply incorporating neighboring agents' policies might not be sophisticated enough to accurately model and adapt to the intricacies among agents.

\vspace{-10pt}
\subsection{Performance of the Quantization-based MACACC}
In this subsection, we evaluate the effectiveness of the proposed quantization-based communication protocol with different quantization resolutions. As shown in Figure~\ref {fig:train_reward_q}, in the less complex ``Catchup'' scenario, minor quantization appears to improve control performance. This could be attributed to the fact that the quantization process introduces a level of randomness during the training phase, thereby fostering improved exploration by the agents, as discussed in \cite{plappert2017parameter}. Conversely, in the more challenging ``Slowdown'' scenario, the impact of quantization results in a more significant performance degradation relative to the ``Catchup'' scenario. Nonetheless, even with extremely quantized communication, such as QMACACC ($1$), our proposed approach continues to surpass the performance of the robust baseline method, FPrint.

Figure~\ref{fig:comm_efficient} presents the number of bits required for each communicated parameter as well as the corresponding test performance at varying quantization resolutions. For better visualization, these values are normalized with corresponding maximum values. Within the ``Catchup'' scenario, QMACACC ($1$) manages to achieve 98.63\% of the control performance achieved by the non-quantized version, i.e., QMACACC ($0$), while only requiring 12.5\% of the communicated bits. However, in the ``Slowdown'' scenario, QMACACC ($1$) can only realize 64.64\% of the control performance of the non-quantized version, i.e., QMACACC ($0$). This underscores a trade-off between the benefits of enhanced communication efficiency brought about by quantization and the associated diminution in control performance.

\vspace{-10pt}
\subsection{Impact of Platoon Size} \label{sec:5-4}
In this subsection, we explore how variations in platoon sizes affect the performance of our model. The normalized training curves comparison among MACACC, QMACACC (1), and the top-performing baseline methods under different platoon sizes (i.e., $\mathcal{V} \in {2, 8, 12}$) in the two CACC scenarios is illustrated in Figure~\ref{fig:train_reward_psize}. Our current experimental design, which encompasses scenarios with up to 12 AVs, has been carefully chosen to reflect practical, real-world traffic scenarios. These experiments are intended to simulate a range of traffic densities, from sparse (i.e., 2 AVs) to moderately dense (i.e., 8 and 12 AVs) conditions, thereby providing a comprehensive overview of the framework's scalability and effectiveness across different urban traffic situations. Given the anticipated increase in CAVs within future urban environments, we plan to extend our investigation into the framework's scalability aimed at scenarios with higher CAV densities in forthcoming research efforts.

As the platoon size increases, the performance of all algorithms decreases in both CACC scenarios. That is attributed to the increased complexities and intricacies associated with managing and coordinating a larger number of agents. As the platoon size increases, the algorithms must deal with more sophisticated inter-agent dynamics, thereby extending the computation time and potentially slowing down the convergence toward an optimal policy. Furthermore, minor quantization errors may accumulate over time or across a platoon, leading to significant deviations from ideal behavior, a phenomenon that becomes increasingly pronounced with larger platoon sizes.
Despite the varying platoon sizes, all the algorithms achieve comparable performance in the ``Catchup'' scenario. However, in the more challenging ``Slowdown'' environment, MACACC ($n={0, 1}$) consistently outperforms the baseline method (i.e., FPrint) under different platoon sizes, showing the impressive scalability of our proposed approach.

\vspace{-10pt}
\subsection{Robustness to Unseen Scenarios} \label{sec:5-5}
In this subsection, we assess MACACC's performance in new scenarios. Although the algorithms were trained as outlined in Section~\ref{sec:2cacc}, they were evaluated in distinct scenarios. In the ``Catchup'' environment, the platoon leader (PL) is initially set with states $v_{1, 0}= v^*_t$ and $h_{1, 0} = a \cdot h^*_t$, for $a \in [1.5, 2.5]$, and subsequently tested with $a \in [2.5, 3.5]$. In the ``Slowdown'' scenario, every vehicle ($i = 1, \cdots, \mathcal{V}$) starts with velocities $v_{i, 0} = b \cdot v^*_t$ and $h_{i, 0} = h^*_t$, where $b \in [1.5, 2.5]$, and is tested on $b \in [0.5, 1.5]$. Table~\ref{tab:iii} shows the execution results in these unexplored scenarios. It is clear that MACACC consistently surpasses baseline methods in both settings. However, MACACC (1) shows slightly inferior performance compared to MACACC, attributable to quantization.

\begin{table}[!ht]
\renewcommand{\arraystretch}{1.4}
\centering
\caption{Execution performance comparison on unseen scenarios.}
\label{tab:iii}
\begin{tabular}{c|c|c|c|c}
\hline
\textbf{Scenario Name} & \textbf{ConseNet} & \textbf{FPrint} & \textbf{MACACC (1)} & \textbf{MACACC} \\ \hline \hline
Catchup & -569.12 & -982.58 & -173.32 & -167.34 \\ \hline
Slowdown & -621.19 & -560.22 & -224.41 & -153.22 \\ \hline
\end{tabular}
\vspace{-5pt}
\end{table}

Table~\ref{tab:iv} provides a detailed evaluation of the performance of MARL controllers, focusing on temporal average metrics within scenarios not encountered during training. It is clearly demonstrated that MACACC consistently outperforms the baseline methods, achieving a remarkable zero collision rate in both ``Catchup'' and ``Slowdown'' scenarios. This underscores MACACC's superior ability to manage complex traffic situations effectively. However, when examining MACACC (1), a noticeable dip in performance is observed in the ``Slowdown'' scenario. This highlights a critical trade-off: while quantization promotes communication efficiency among agents, it simultaneously can lead to a reduction in control performance. This balance between the benefits of streamlined communication and the potential impact on decision-making underscores the nuanced challenges inherent in optimizing MARL systems for complex systems.

\begin{table}[!ht]
\renewcommand{\arraystretch}{1.4}
\centering
\caption{Performance of MARL controllers in terms of temporal average metrics on unseen scenarios: ``Catchup'' (above) and ``Slowdown'' (below).}
\label{tab:iv}
\begin{tabular}{l|c|c|c|c}
\hline
\textbf{Metrics}       & \textbf{ConseNet} & \textbf{FPrint} & \textbf{MACACC(1)} & \textbf{MACACC} \\ \hline \hline
avg. headway {[}m{]}   & 14.82             & 12.51           & 20.8               & 20.5            \\ 
avg velocity {[}m/s{]} & 10.82             & 9.28            & 15.65              & 15.7            \\ 
collision number          & 15                & 20              & 0                  & 0               \\ \hline
avg. headway {[}m{]}   & 17.92             & 16.34           & 18.99              & 20.54           \\ 
avg velocity {[}m/s{]} & 13.02             & 12.3            & 14.28              & 14.8            \\ 
collision number           & 10                & 6               & 2                  & 0               \\ \hline
\end{tabular}
\vspace{-10pt}
\end{table}

\vspace{-5pt}
\section{Conclusion}\label{sec:6}
Cooperative adaptive cruise control (CACC), has been recognized for its capability to increase road efficiency, alleviate traffic congestion, and reduce both energy consumption and exhaust emissions. In this paper, we have addressed the CACC problem by formulating it as a fully decentralized MARL problem. (1) This novel approach eliminated the need for a centralized controller during both training and execution, thereby enhancing the system's scalability and robustness; (2) an innovative quantization-based communication protocol was introduced to significantly enhance the communication efficiency among the agents; (3) comprehensive experiments were conducted showing that our approach outperformed several state-of-the-art MARL algorithms. The results demonstrated that our approach can provide superior control performance and communication efficiency; (4) as detailed in the appendix, our proposed framework's applicability extends beyond CACC, showing promise for broader intelligent transportation systems characterized by intricate action and state spaces. These findings underscore the potential of our fully decentralized MARL and quantization-based communication protocol as a robust and effective solution for real-world MARL problems.

In this paper, we employed the optimal velocity model (OVM) to emulate certain aspects of traffic flow behaviors due to its simplicity and efficacy. However, it is worth noting that OVM oversimplifies some intricate nature of human driving behaviors, and its performance may degrade at high traffic densities where interactions between vehicles become more complex. As a result, future research endeavors will focus on the integration of more comprehensive human driver models and vehicle dynamics to improve simulation accuracy and stability. Furthermore, to eliminate the need for a training process, we plan to explore the best response dynamics approach, as proposed in \cite{dai2021towards}, where agents communicate their actions using the QMACACC protocol. Also, the issue of energy management within cooperative scenarios will be investigated, drawing upon insights from recent studies such as \cite{hua2023energy}. Additionally, drawing inspiration from the novel research by \cite{xia2023secure} showing the robustness of the cooperative localization system within a consensus framework and incorporating a novel cyber attack detection algorithm, our future study aims to explore the resilience of MARL in the face of potential cyber-attacks and uncertainties.

\appendix
In the Appendix, we evaluate our proposed communication-efficient MARL framework on a more complex traffic scenario: adaptive traffic signal control (ATSC). The goal of ATSC is to dynamically modify traffic signal phases to alleviate traffic congestion, leveraging real-time traffic data \cite{chu2020multiagent}. We adopt the ATSC scenario, a 5$\times$5 synthetic traffic grid, using standard microscopic traffic simulator SUMO~\cite{krajzewicz2012recent, chu2020multiagent}.

\vspace{-5pt}
\subsection{Experimental settings}
Similar to previous studies \cite{chu2019multi, chu2020multiagent}, the ATSC scenarios within our framework are designed to replicate peak-hour traffic conditions. A 5-second control interval is strategically implemented to prevent too frequent changes in traffic light signals, a decision that accommodates the delay associated with RL control mechanisms and the anticipated reaction time of drivers. As a result, each Markov Decision Process (MDP) step corresponds to a real-time duration of 5 seconds, with the total episode horizon extending to 720 steps to cover a comprehensive simulation interval. To further ensure road safety and provide adequate reaction time for drivers, a 2-second yellow light phase is introduced before the transition to red lights. For readers interested in a detailed understanding of the ATSC settings, further insights can be found in the works of \cite{chu2019multi, chu2020multiagent}.

\begin{figure}[!ht]
  \centering
  \subfloat[Synthetic traffic grid.]{%
    \includegraphics[width=0.36\textwidth]{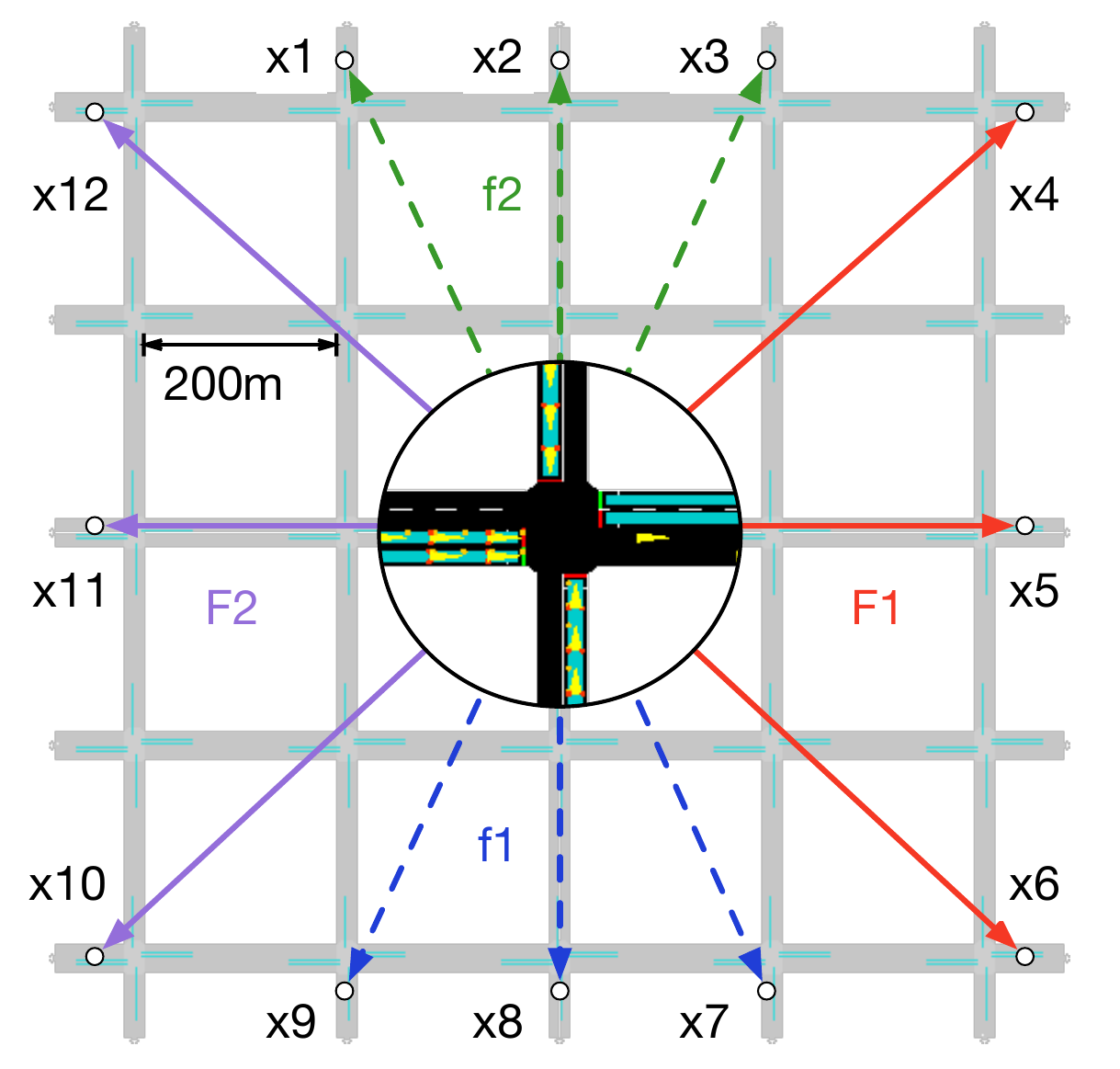}%
    \label{fig:atsc1}%
  }
  \hfill
  \subfloat[Traffic flows within the grid.]{%
    \includegraphics[width=0.4\textwidth]{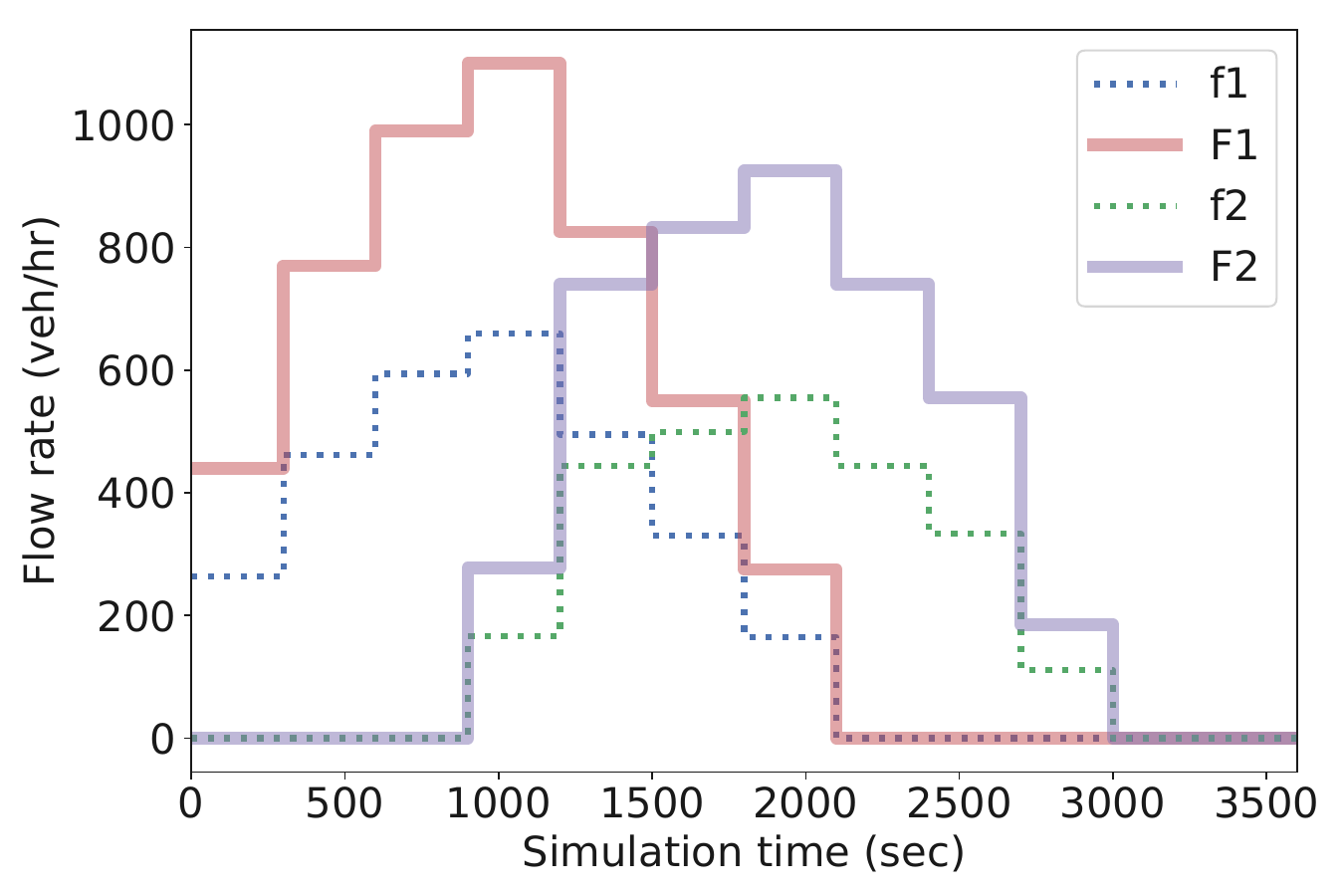}%
    \label{fig:atsc2}%
  }
  \caption{ATSC Scenario Illustrations: (a) Synthetic traffic grid indicating major (solid arrows) and minor (dotted arrows) flows; (b) Simulation of dynamic traffic flows over time within the grid. Adapted from~\cite{chu2019multi, chu2020multiagent}.}
  \label{fig:atsc}
\end{figure}

Fig.~\ref{fig:atsc1} illustrates the structured traffic grid, including arterial streets with two lanes and a 20 m/s speed limit alongside avenues that feature a single lane and an 11 m/s speed limit. To model the intricacies of peak-hour traffic, we employ four sets of dynamic traffic flows, capturing both the loading and recovering phases.  In the initial stage, we introduce three primary flows, denoted as $F_1$, originating and terminating at pairs $x_{10}$-$x_{4}$, $x_{11}$-$x_{5}$, and $x_{12}$-$x_{6}$. Concurrently, three secondary flows, labeled $f_1$, emerge with origin-destination pairs $x_{1}$-$x_{7}$, $x_{2}$-$x_{8}$, and $x_{3}$-$x_{9}$. Following a 15-minute interval, both $F_1$ and $f_1$ begin to wane, giving rise to their respective counterflows, $F_2$ and $f_2$, as illustrated in Fig.~\ref{fig:atsc2}. It is important to note that these flows merely establish overarching demand; the specific trajectory of each vehicle is determined through random generation. The grid operates under uniform conditions, with all agents sharing an identical set of possible actions, defined by five predetermined signal phases \cite{chu2019multi, chu2020multiagent}.

\begin{figure*}[!ht]
  \centering
  \includegraphics[width=0.85\textwidth]{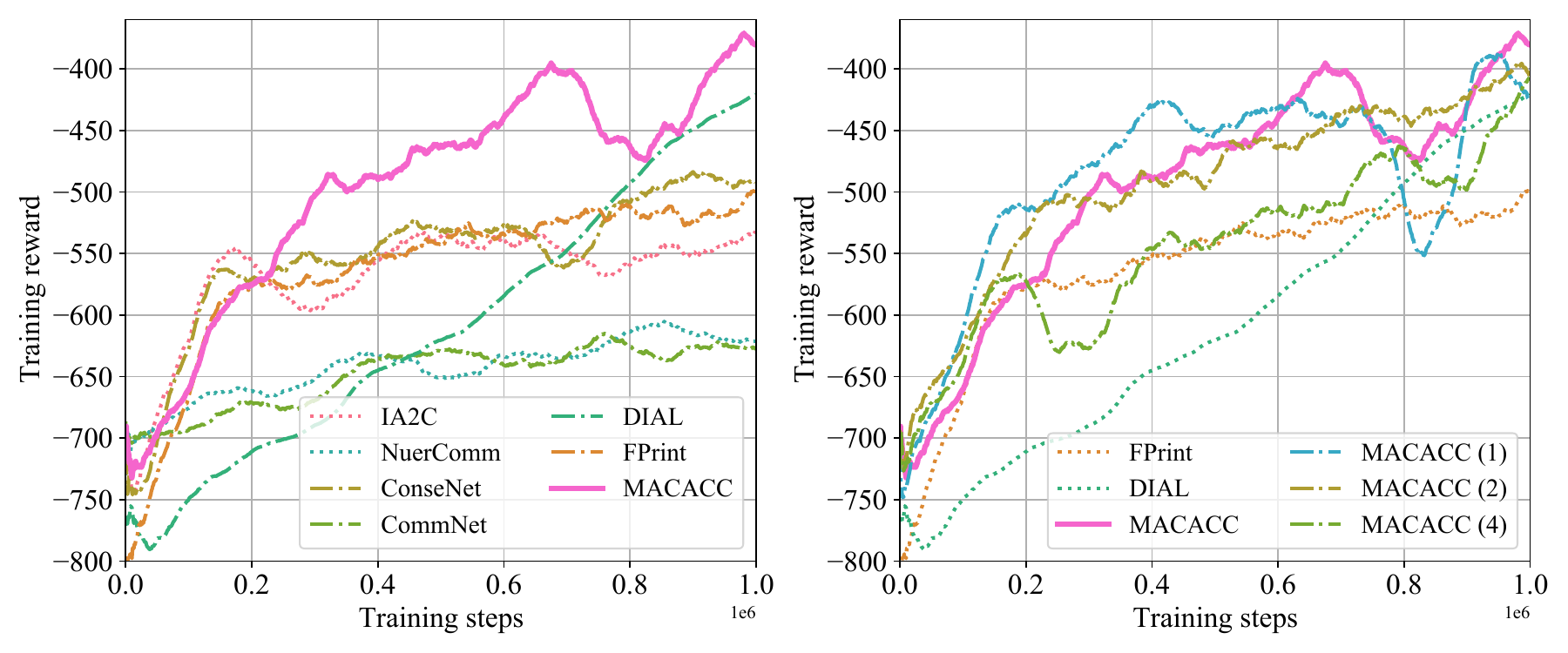}
  \caption{Training curves comparison between the proposed MARL variant (MACACC ($n$)) and 6 state-of-the-art MARL benchmarks on adaptive traffic signal control (ATSC).}
  \label{fig:train_reward_atsc}
  \vspace{-10pt}
\end{figure*}

\vspace{-5pt}
\subsection{MDP settings}
We conceptualize the ATSC problem as a model-free multi-agent network, treating each intersection (totaling 25 intersections) as a controllable agent. This collective dynamic framework is structured as a Markov Decision Process (MDP), denoted by $\mathcal{M}_A = (\mathbf{A}, \mathbf{S}, \mathbf{R}\}, \mathbf{T})$ \cite{chu2019multi, chu2020multiagent}:
\begin{itemize}
    \item \textbf{Action Space $\mathbf{A}$}: The choice of action at each intersection involves selecting a possible phase or a combination of red and green signals for the traffic lights. For each intersection, we consider 5 possible phases and the total action space expands to $5^{25}$.
    \item \textbf{State Space $\mathbf{S}$}: The state $s_t$ for each agent includes local traffic conditions as $[\{\text{wait}_t[p], \text{wave}_t[p]\}]$, where $p$ represents each incoming lane to intersection $i$. $\text{wait}_t[p]$ captures the cumulative waiting time of the first vehicle in lane $p$, and $\text{wave}_t[p]\}]$ quantifies the total vehicle count approaching each incoming lane within a 50m radius of the intersection. The state space for each agent ranges from 36 to 60 states, resulting in a total state space scope between $30^{25}$ to $60^{25}$. 
    \item \textbf{Reward Function $\mathbf{R}$}: The reward function is designed to facilitate the RL agents' training and is defined as follows:
    \begin{equation}
        r_t = - \sum (\text{queue}_{t+ \Delta_t}[p] + a \cdot \text{wait}_{t+ \Delta_t}[p]).
    \end{equation}
    where $a$ is a balancing coefficient, and $\text{queue}_{t+ \Delta_t}[p]$ indicates the queue length along each lane approaching the intersection. It is noteworthy that the reward is calculated post-decision, hence both $queue$ and $wait$ are evaluated at the future time $t+ \Delta_t$.
    \item  \textbf{Transition Probability $\mathbf{T}$}: Given the model-free nature of our approach, we proceed without preconceived assumptions regarding the system's dynamics.
\end{itemize}

\subsection{Experimental results}
Figure~\ref{fig:train_reward_atsc} shows training curves comparison between the proposed MARL variant (MACACC ($n$)) and six state-of-the-art MARL benchmarks on the ATSC problem. Note that we focus only on the training reward as a metric to calculate overall performance. A more comprehensive evaluation will be addressed in future work. It is evident that MACACC demonstrates the quickest convergence speed and obtains the highest training reward compared to other state-of-the-art MARL benchmarks, underscoring the effectiveness of the MACACC framework in learning and optimizing traffic signal control strategies more efficiently. Following closely, DIAL also delivers commendable performance, securing the second-highest training reward. Interestingly, FPrint and ConseNet, known for their effectiveness in the CACC problem, exhibit satisfactory performance in the ATSC context as well. In contrast, CommNet and NeurComm struggle with the control tasks, showing low training rewards and underperforming in comparison to the IA2C approach. This observation suggests varying degrees of adaptability and effectiveness among MARL solutions when applied to the complexities of the ATSC problem. Figure~\ref{fig:train_reward_atsc} also shows that even with extremely quantized communication, such as QMACACC (2), our proposed approach consistently surpasses the performance of the robust baseline methods, DIAL and FPrint. Nonetheless, adopting an even more stringent quantization strategy, i.e., QMACACC (4), leads to a noticeable decline in performance. This indicates that while quantization enhances communication efficiency, there is a critical threshold beyond which the reduction in communication detail adversely affects the overall system performance, underscoring the importance of balancing communication efficiency and information sharing.

\bibliography{ref_tiv}

\begin{thebibliography}{10}
\providecommand{\url}[1]{#1}
\csname url@samestyle\endcsname
\providecommand{\newblock}{\relax}
\providecommand{\bibinfo}[2]{#2}
\providecommand{\BIBentrySTDinterwordspacing}{\spaceskip=0pt\relax}
\providecommand{\BIBentryALTinterwordstretchfactor}{4}
\providecommand{\BIBentryALTinterwordspacing}{\spaceskip=\fontdimen2\font plus
\BIBentryALTinterwordstretchfactor\fontdimen3\font minus \fontdimen4\font\relax}
\providecommand{\BIBforeignlanguage}[2]{{%
\expandafter\ifx\csname l@#1\endcsname\relax
\typeout{** WARNING: IEEEtran.bst: No hyphenation pattern has been}%
\typeout{** loaded for the language `#1'. Using the pattern for}%
\typeout{** the default language instead.}%
\else
\language=\csname l@#1\endcsname
\fi
#2}}
\providecommand{\BIBdecl}{\relax}
\BIBdecl

\bibitem{SOS}
L.~Chen, Y.~Li, C.~Huang \emph{et~al.}, ``Milestones in autonomous driving and intelligent vehicles: Survey of surveys,'' \emph{IEEE Transactions on Intelligent Vehicles}, vol.~8, no.~2, pp. 1046--1056, 2023.

\bibitem{han2023secure}
J.~Han, Z.~Ju, X.~Chen, M.~Yang, H.~Zhang, and R.~Huai, ``Secure operations of connected and autonomous vehicles,'' \emph{IEEE Transactions on Intelligent Vehicles}, 2023.

\bibitem{Planning}
S.~Teng, X.~Hu, P.~Deng, B.~Li \emph{et~al.}, ``Motion planning for autonomous driving: The state of the art and future perspectives,'' \emph{IEEE Transactions on Intelligent Vehicles}, vol.~8, no.~6, pp. 3692--3711, 2023.

\bibitem{cao2022future}
D.~Cao, X.~Wang, L.~Li, C.~Lv, X.~Na, Y.~Xing, X.~Li, Y.~Li, Y.~Chen, and F.-Y. Wang, ``Future directions of intelligent vehicles: Potentials, possibilities, and perspectives,'' \emph{IEEE Transactions on Intelligent Vehicles}, vol.~7, no.~1, pp. 7--10, 2022.

\bibitem{zhang2023emerging}
H.~Zhang, J.~Guo, G.~Luo, L.~Li, X.~Na, X.~Wang, S.~Teng, S.~Ma, and Y.~Li, ``Emerging trends in intelligent vehicles: The ieee tiv perspective,'' \emph{IEEE Transactions on Intelligent Vehicles}, 2023.

\bibitem{kazemi2018learning}
H.~Kazemi, H.~N. Mahjoub, A.~Tahmasbi-Sarvestani, and Y.~P. Fallah, ``A learning-based stochastic mpc design for cooperative adaptive cruise control to handle interfering vehicles,'' \emph{IEEE Transactions on Intelligent Vehicles}, vol.~3, no.~3, pp. 266--275, 2018.

\bibitem{wu2023review}
C.~Wu, Z.~Cai, Y.~He, and X.~Lu, ``A review of vehicle group intelligence in a connected environment,'' \emph{IEEE Transactions on Intelligent Vehicles}, 2023.

\bibitem{bolduc2019multimodel}
A.~P. Bolduc, L.~Guo, and Y.~Jia, ``Multimodel approach to personalized autonomous adaptive cruise control,'' \emph{IEEE Transactions on Intelligent Vehicles}, vol.~4, no.~2, pp. 321--330, 2019.

\bibitem{wen2018cooperative}
S.~Wen, G.~Guo, B.~Chen, and X.~Gao, ``Cooperative adaptive cruise control of vehicles using a resource-efficient communication mechanism,'' \emph{IEEE Transactions on Intelligent Vehicles}, vol.~4, no.~1, pp. 127--140, 2018.

\bibitem{lin2020comparison}
Y.~Lin, J.~McPhee, and N.~L. Azad, ``Comparison of deep reinforcement learning and model predictive control for adaptive cruise control,'' \emph{IEEE Transactions on Intelligent Vehicles}, vol.~6, no.~2, pp. 221--231, 2020.

\bibitem{wang2020online}
Y.~Wang, G.~Gunter, M.~Nice, M.~L. Delle~Monache, and D.~B. Work, ``Online parameter estimation methods for adaptive cruise control systems,'' \emph{IEEE Transactions on Intelligent Vehicles}, vol.~6, no.~2, pp. 288--298, 2020.

\bibitem{hu2023cacc}
J.~Hu, S.~Sun, J.~Lai, S.~Wang, Z.~Chen, and T.~Liu, ``Cacc simulation platform designed for urban scenes,'' \emph{IEEE Transactions on Intelligent Vehicles}, 2023.

\bibitem{althoff2020provably}
M.~Althoff, S.~Maierhofer, and C.~Pek, ``Provably-correct and comfortable adaptive cruise control,'' \emph{IEEE Transactions on Intelligent Vehicles}, vol.~6, no.~1, pp. 159--174, 2020.

\bibitem{massera2017safe}
C.~Massera~Filho, M.~H. Terra, and D.~F. Wolf, ``Safe optimization of highway traffic with robust model predictive control-based cooperative adaptive cruise control,'' \emph{IEEE Transactions on Intelligent Transportation Systems}, vol.~18, no.~11, pp. 3193--3203, 2017.

\bibitem{gao2016data}
W.~Gao, Z.-P. Jiang, and K.~Ozbay, ``Data-driven adaptive optimal control of connected vehicles,'' \emph{IEEE Transactions on Intelligent Transportation Systems}, vol.~18, no.~5, pp. 1122--1133, 2016.

\bibitem{al2017feedforward}
A.~M. Al-Jhayyish and K.~W. Schmidt, ``Feedforward strategies for cooperative adaptive cruise control in heterogeneous vehicle strings,'' \emph{IEEE Transactions on Intelligent Transportation Systems}, vol.~19, no.~1, pp. 113--122, 2017.

\bibitem{wu2018stabilizing}
C.~Wu, A.~M. Bayen, and A.~Mehta, ``Stabilizing traffic with autonomous vehicles,'' in \emph{Proceedings of 2018 IEEE International Conference on Robotics and Automation}, 2018, pp. 6012--6018.

\bibitem{gong2019cooperative}
S.~Gong, A.~Zhou, and S.~Peeta, ``Cooperative adaptive cruise control for a platoon of connected and autonomous vehicles considering dynamic information flow topology,'' \emph{Transportation research record}, vol. 2673, no.~10, pp. 185--198, 2019.

\bibitem{wang2018infrastructure}
M.~Wang, ``Infrastructure assisted adaptive driving to stabilise heterogeneous vehicle strings,'' \emph{Transportation Research Part C: Emerging Technologies}, vol.~91, pp. 276--295, 2018.

\bibitem{feng2019string}
S.~Feng, Y.~Zhang, S.~E. Li, Z.~Cao, H.~X. Liu, and L.~Li, ``String stability for vehicular platoon control: Definitions and analysis methods,'' \emph{Annual Reviews in Control}, vol.~47, pp. 81--97, 2019.

\bibitem{jin2014dynamics}
I.~G. Jin and G.~Orosz, ``Dynamics of connected vehicle systems with delayed acceleration feedback,'' \emph{Transportation Research Part C: Emerging Technologies}, vol.~46, pp. 46--64, 2014.

\bibitem{chu2019model}
T.~Chu and U.~Kalabi{\'c}, ``Model-based deep reinforcement learning for cacc in mixed-autonomy vehicle platoon,'' in \emph{Proceedings of 2019 IEEE 58th Conference on Decision and Control}, 2019, pp. 4079--4084.

\bibitem{lei2022deep}
L.~Lei, T.~Liu, K.~Zheng, and L.~Hanzo, ``Deep reinforcement learning aided platoon control relying on v2x information,'' \emph{IEEE Transactions on Vehicular Technology}, vol.~71, no.~6, pp. 5811--5826, 2022.

\bibitem{jiang2022reinforcement}
L.~Jiang, Y.~Xie, N.~G. Evans, X.~Wen, T.~Li, and D.~Chen, ``Reinforcement learning based cooperative longitudinal control for reducing traffic oscillations and improving platoon stability,'' \emph{Transportation Research Part C: Emerging Technologies}, vol. 141, p. 103744, 2022.

\bibitem{zhu2022joint}
H.~Zhu, Y.~Zhou, X.~Luo, and H.~Zhou, ``Joint control of power, beamwidth, and spacing for platoon-based vehicular cyber-physical systems,'' \emph{IEEE Transactions on Vehicular Technology}, vol.~71, no.~8, pp. 8615--8629, 2022.

\bibitem{wang2022design}
Z.~Wang, S.~Jin, L.~Liu, C.~Fang, M.~Li, and S.~Guo, ``Design of intelligent connected cruise control with vehicle-to-vehicle communication delays,'' \emph{IEEE Transactions on Vehicular Technology}, vol.~71, no.~8, pp. 9011--9025, 2022.

\bibitem{liu2022autonomous}
T.~Liu, L.~Lei, K.~Zheng, and K.~Zhang, ``Autonomous platoon control with integrated deep reinforcement learning and dynamic programming,'' \emph{IEEE Internet of Things Journal}, vol.~10, no.~6, pp. 5476--5489, 2022.

\bibitem{li2023anti}
M.~Li, Z.~Li, S.~Wang, and B.~Wang, ``Anti-disturbance self-supervised reinforcement learning for perturbed car-following system,'' \emph{IEEE Transactions on Vehicular Technology}, 2023.

\bibitem{haarnoja2018soft}
T.~Haarnoja, A.~Zhou, P.~Abbeel, and S.~Levine, ``Soft actor-critic: Off-policy maximum entropy deep reinforcement learning with a stochastic actor,'' in \emph{Proceedings of International conference on machine learning}.\hskip 1em plus 0.5em minus 0.4em\relax PMLR, 2018, pp. 1861--1870.

\bibitem{lillicrap2015continuous}
T.~P. Lillicrap, J.~J. Hunt, A.~Pritzel, N.~Heess, T.~Erez, Y.~Tassa, D.~Silver, and D.~Wierstra, ``Continuous control with deep reinforcement learning,'' \emph{arXiv preprint arXiv:1509.02971}, 2015.

\bibitem{desjardins2011cooperative}
C.~Desjardins and B.~Chaib-Draa, ``Cooperative adaptive cruise control: A reinforcement learning approach,'' \emph{IEEE Transactions on intelligent transportation systems}, vol.~12, no.~4, pp. 1248--1260, 2011.

\bibitem{jia2016enhanced}
D.~Jia and D.~Ngoduy, ``Enhanced cooperative car-following traffic model with the combination of v2v and v2i communication,'' \emph{Transportation Research Part B: Methodological}, vol.~90, pp. 172--191, 2016.

\bibitem{chu2020multiagent}
\BIBentryALTinterwordspacing
T.~Chu, S.~Chinchali, and S.~Katti, ``Multi-agent reinforcement learning for networked system control,'' in \emph{Proceedings of International Conference on Learning Representations}, 2020. [Online]. Available: \url{https://openreview.net/forum?id=Syx7A3NFvH}
\BIBentrySTDinterwordspacing

\bibitem{peake2020multi}
A.~Peake, J.~McCalmon, B.~Raiford, T.~Liu, and S.~Alqahtani, ``Multi-agent reinforcement learning for cooperative adaptive cruise control,'' in \emph{Proceedings of 2020 IEEE 32nd International Conference on Tools with Artificial Intelligence}, 2020, pp. 15--22.

\bibitem{raja2022blockchain}
G.~Raja, K.~Kottursamy, K.~Dev, R.~Narayanan, A.~Raja, and K.~B.~V. Karthik, ``Blockchain-integrated multiagent deep reinforcement learning for securing cooperative adaptive cruise control,'' \emph{IEEE transactions on intelligent transportation systems}, vol.~23, no.~7, pp. 9630--9639, 2022.

\bibitem{hua2023energy}
\BIBentryALTinterwordspacing
M.~Hua, C.~Zhang, F.~Zhang, Z.~Li, X.~Yu, H.~Xu, and Q.~Zhou, ``Energy management of multi-mode plug-in hybrid electric vehicle using multi-agent deep reinforcement learning,'' \emph{Applied Energy}, vol. 348, p. 121526, 2023. [Online]. Available: \url{https://www.sciencedirect.com/science/article/pii/S0306261923008905}
\BIBentrySTDinterwordspacing

\bibitem{lowe2017multi}
R.~Lowe, Y.~I. Wu, A.~Tamar, J.~Harb, O.~Pieter~Abbeel, and I.~Mordatch, ``Multi-agent actor-critic for mixed cooperative-competitive environments,'' \emph{Advances in neural information processing systems}, vol.~30, 2017.

\bibitem{zhang2018fully}
K.~Zhang, Z.~Yang, H.~Liu, T.~Zhang, and T.~Basar, ``Fully decentralized multi-agent reinforcement learning with networked agents,'' in \emph{Proceedings of International Conference on Machine Learning}.\hskip 1em plus 0.5em minus 0.4em\relax PMLR, 2018, pp. 5872--5881.

\bibitem{zhang2021multi}
K.~Zhang, Z.~Yang, and T.~Ba{\c{s}}ar, ``Multi-agent reinforcement learning: A selective overview of theories and algorithms,'' \emph{Handbook of reinforcement learning and control}, pp. 321--384, 2021.

\bibitem{mnih2015human}
V.~Mnih, K.~Kavukcuoglu, D.~Silver, A.~A. Rusu, J.~Veness, M.~G. Bellemare, A.~Graves, M.~Riedmiller, A.~K. Fidjeland, G.~Ostrovski \emph{et~al.}, ``Human-level control through deep reinforcement learning,'' \emph{nature}, vol. 518, no. 7540, pp. 529--533, 2015.

\bibitem{mnih2016asynchronous}
V.~Mnih, A.~P. Badia, M.~Mirza, A.~Graves, T.~Lillicrap, T.~Harley, D.~Silver, and K.~Kavukcuoglu, ``Asynchronous methods for deep reinforcement learning,'' in \emph{International conference on machine learning}.\hskip 1em plus 0.5em minus 0.4em\relax PMLR, 2016, pp. 1928--1937.

\bibitem{watkins1992q}
C.~J. Watkins and P.~Dayan, ``Q-learning,'' \emph{Machine learning}, vol.~8, pp. 279--292, 1992.

\bibitem{szepesvari2022algorithms}
C.~Szepesv{\'a}ri, \emph{Algorithms for reinforcement learning}.\hskip 1em plus 0.5em minus 0.4em\relax Springer Nature, 2022.

\bibitem{chu2019multi}
T.~Chu, J.~Wang, L.~Codec{\`a}, and Z.~Li, ``Multi-agent deep reinforcement learning for large-scale traffic signal control,'' \emph{IEEE Transactions on Intelligent Transportation Systems}, vol.~21, no.~3, pp. 1086--1095, 2019.

\bibitem{chen2023deep}
D.~Chen, M.~R. Hajidavalloo, Z.~Li, K.~Chen, Y.~Wang, L.~Jiang, and Y.~Wang, ``Deep multi-agent reinforcement learning for highway on-ramp merging in mixed traffic,'' \emph{IEEE Transactions on Intelligent Transportation Systems}, 2023.

\bibitem{tan1993multi}
M.~Tan, ``Multi-agent reinforcement learning: Independent vs. cooperative agents,'' in \emph{Proceedings of the Tenth International Conference on Machine Learning}, 1993, pp. 330--337.

\bibitem{foerster2017stabilising}
J.~Foerster, N.~Nardelli, G.~Farquhar, T.~Afouras, P.~H. Torr, P.~Kohli, and S.~Whiteson, ``Stabilising experience replay for deep multi-agent reinforcement learning,'' in \emph{Proceedings of International conference on machine learning}.\hskip 1em plus 0.5em minus 0.4em\relax PMLR, 2017, pp. 1146--1155.

\bibitem{foerster2016learning}
J.~Foerster, I.~A. Assael, N.~De~Freitas, and S.~Whiteson, ``Learning to communicate with deep multi-agent reinforcement learning,'' \emph{Advances in neural information processing systems}, vol.~29, 2016.

\bibitem{sukhbaatar2016learning}
S.~Sukhbaatar, R.~Fergus \emph{et~al.}, ``Learning multiagent communication with backpropagation,'' \emph{Advances in neural information processing systems}, vol.~29, 2016.

\bibitem{chen2021powernet}
D.~Chen, K.~Chen, Z.~Li, T.~Chu, R.~Yao, F.~Qiu, and K.~Lin, ``Powernet: Multi-agent deep reinforcement learning for scalable powergrid control,'' \emph{IEEE Transactions on Power Systems}, vol.~37, no.~2, pp. 1007--1017, 2021.

\bibitem{bando1995dynamical}
M.~Bando, K.~Hasebe, A.~Nakayama, A.~Shibata, and Y.~Sugiyama, ``Dynamical model of traffic congestion and numerical simulation,'' \emph{Physical review E}, vol.~51, no.~2, p. 1035, 1995.

\bibitem{alistarh2017qsgd}
D.~Alistarh, D.~Grubic, J.~Li, R.~Tomioka, and M.~Vojnovic, ``{QSGD}: Communication-efficient {SGD} via gradient quantization and encoding,'' in \emph{Proceedings of Advances in Neural Information Processing Systems}, 2017, pp. 1709--1720.

\bibitem{wang2022quantization}
Y.~Wang and T.~Ba{\c{s}}ar, ``Quantization enabled privacy protection in decentralized stochastic optimization,'' \emph{IEEE Transactions on Automatic Control}, vol.~68, no.~7, pp. 4038--4052, 2023.

\bibitem{saxe2013exact}
A.~M. Saxe, J.~L. McClelland, and S.~Ganguli, ``Exact solutions to the nonlinear dynamics of learning in deep linear neural networks,'' \emph{arXiv preprint arXiv:1312.6120}, 2013.

\bibitem{plappert2017parameter}
M.~Plappert, R.~Houthooft, P.~Dhariwal, S.~Sidor, R.~Y. Chen, X.~Chen, T.~Asfour, P.~Abbeel, and M.~Andrychowicz, ``Parameter space noise for exploration,'' \emph{arXiv preprint arXiv:1706.01905}, 2017.

\bibitem{dai2021towards}
Q.~Dai, X.~Xu, W.~Guo, S.~Huang, and D.~Filev, ``Towards a systematic computational framework for modeling multi-agent decision-making at micro level for smart vehicles in a smart world,'' \emph{Robotics and Autonomous Systems}, vol. 144, p. 103859, 2021.

\bibitem{xia2023secure}
X.~Xia, R.~Xu, and J.~Ma, ``Secure cooperative localization for connected automated vehicles based on consensus,'' \emph{IEEE Sensors Journal}, 2023.

\bibitem{krajzewicz2012recent}
D.~Krajzewicz, J.~Erdmann, M.~Behrisch, and L.~Bieker, ``Recent development and applications of sumo-simulation of urban mobility,'' \emph{International journal on advances in systems and measurements}, vol.~5, no. 3\&4, 2012.

\end{thebibliography}
\bibliographystyle{IEEEtran}

\end{document}